\documentclass[preprint,12pt,sort&compress]{elsarticle}
\journal{arXiv}
\usepackage{color}
\usepackage{graphicx}
\usepackage{amssymb}
\usepackage{tcolorbox}
\tcbuselibrary{skins, breakable}
\usepackage{amsmath}
\usepackage{aeguill}
\usepackage{cases}
\usepackage{fancybox}
\usepackage{version}
\usepackage{enumitem}
\usepackage{mathtools}
\usepackage{mathrsfs}
\usepackage{yhmath}
\usepackage{stmaryrd}
\usepackage{subfigure}
\usepackage[Symbol]{upgreek}
\usepackage{mathrsfs}
\usepackage{bm}
\usepackage{url}
\usepackage{float}
\usepackage{caption}
\usepackage{booktabs}
\usepackage{siunitx}
\usepackage{amsmath}    
\usepackage{array}        
\usepackage{booktabs}     
\usepackage{caption}      
\usepackage{geometry}     
\usepackage[colorlinks,
linkcolor=black,
anchorcolor=black,
citecolor=black
]{hyperref}
\usepackage[normalem]{ulem}

\begin{document}
	\captionsetup[figure]{labelfont={bf},name={Fig.},labelsep=period}
	\begin{frontmatter}
		\author[DLUT]{Wenping Han}
		\author[DLUT]{Bowen Sun}
		\author[DLUT]{Shankun Liu}
		\author[DLUT]{Fei Han\corref{cor}}
		\cortext[cor]{Corresponding author. hanfei8172@126.com (Fei Han)}
		\address[DLUT]{State Key Laboratory of Structural Analysis, Optimization and CAE Software for Industrial Equipment, Department of Engineering Mechanics, Dalian University of Technology, Dalian, 116023, PR China}
		
		\title{An improved bond-associated peridynamic model and its adaptive coupling with CCM for fracture analysis}
		\begin{abstract}
			
		This paper reformulates the correction factor in the force-state of the bond-associated peridynamic (BAPD) model. The reformulation is established from the strain energy density equivalence between the BAPD model and the classical continuum mechanics (CCM) model at a material point. With the FEM solution taken as the reference, the proposed correction factor improves the accuracy of the BAPD solution in this study. Furthermore, a BAPD-CCM coupled model is developed based on the above energy density equivalence, and a ``Morphing" function is introduced to achieve a smooth transition between the two models. For time integration, explicit schemes are adopted for both quasi-static and dynamic problems. In the spatial discretization, the CCM model is discretized by elements, whereas the BAPD model is discretized by particles. The solution accuracy of the coupled model is validated by comparison with the FEM solution and by evaluating the $L^{2}$ norm, the $H^{1}_{\mathrm{semi}}$, and the energy norm of the displacement error. Two- and three-dimensional numerical examples show that the proposed model has higher computational efficiency. For example, in the Mode I crack propagation problem, its computational cost is reduced by more than 72\% compared with that of the pure BAPD model, and the predicted crack patterns agree with experimental results.

		\end{abstract}
		
		
	\end{frontmatter}
	
	\section{Introduction}
	
	During the service life of engineering structures, crack initiation, propagation, and branching are likely to occur in engineering materials such as metals, geomaterials, and concrete \cite{Anderson2017, Karihaloo1995, Broek1986}. Accurately capturing the full evolution of cracks in numerical simulations is important for engineering practice. 

	To address these challenges, a variety of methods have been developed within the framework of the classical continuum mechanics (CCM) model. Classical fracture mechanics characterizes crack-tip fields by a small set of parameters, such as the stress intensity factor and the energy release rate, and has established a mature system of fracture criteria \cite{Anderson2017,Karihaloo1995,Broek1986}. However, it typically assumes that the crack geometry and propagation path are known in advance. It also relies on a small-scale yielding assumption so that the near-tip fields can be described by only a few parameters; when the fracture process zone is extended, and the cracking pattern becomes highly complex, this few-parameter description becomes limited \cite{Rabczuk2013,Wang2024_MultiCrackReview}. The cohesive zone model (CZM) was proposed by Barenblatt and Dugdale \cite{Barenblatt1962,Dugdale1960}. By prescribing traction-separation relations along potential crack paths, CZM embeds the non-linear fracture zone into the model and can describe different crack patterns and interface debonding in a unified way \cite{Jemblie2017_CZMReview}. However, in conventional finite element (FE) implementations, interface elements are usually pre-inserted along prescribed paths, which forces cracks to propagate along element boundaries and makes the results mesh-dependent \cite{Wcislik2021_CZMSelectedAspects}. Phase-field methods (PFM) introduce a continuous phase-field variable to regularize sharp cracks into finite-width damage bands, so that complex cracks can be captured naturally without explicitly tracking crack surfaces \cite{Bourdin2000,Li2023_PhaseFieldReview}. But the associated energy functionals are highly nonlinear and involve an internal length-scale parameter, which requires sufficiently fine meshes in the fracture zone to resolve the phase-field gradients, leading to substantial computational cost and sensitivity to the chosen length-scale parameters \cite{Li2023_PhaseFieldReview,Cervera2022_XFEM_MixedFEM_PF}.
	
	Unlike the methods discussed above, peridynamics is a non-local solid mechanics theory first proposed by Silling \cite{Silling2000_BBPD} in 2000. In peridynamics, the internal force is written as an integral over the interactions of material points within a finite neighborhood. This integral formulation replaces spatial derivatives with non-local interactions. Therefore, the formulation remains valid even when the displacement field is discontinuous, allowing cracks to be represented naturally. In recent years, peridynamics has been widely employed for the numerical simulation of various complex fracture problems, such as impact and blast loading \cite{Jafaraghaei2022_GlassPD, Zhu2021_PDBlasting, BOWEN_2025}, thermo-mechanical coupling, and electro-chemo-mechanical effects in functional materials \cite{Wang2021_SiThinFilmPD,Wang2022_CoreShellPD}. Meanwhile, peridynamics can be combined with other approaches to provide complementary capabilities, such as PFM and non-local damage theory \cite{Diehl2022_PDPFReview,Oterkus2024_PDRecentReview}.
	
	Nevertheless, despite these successful applications, peridynamic (PD) theory still faces challenges in developing constitutive models suitable for engineering applications. Peridynamic models can be broadly classified into bond-based peridynamic (BBPD) and state-based peridynamic (SBPD) models according to their constitutive formulations. In the BBPD model, the constitutive response of a bond depends only on the deformation of that bond itself, whereas in the SBPD model, it depends on the deformation state of the entire neighborhood \cite{Silling2000_BBPD, Silling2007_StateBased}. However, the equivalent Poisson ratio obtained by the BBPD model is a fixed value (e.g., 1/3 in 2D plane stress), making it difficult to represent typical engineering materials. Fallah et al. \cite{Fallah2020_BBPDConstitutive} and Trageser et al. \cite{Trageser2020_BBPD_Poisson} have systematically analyzed this restriction and proposed modified bond-based constitutive constructions, but these approaches require bond classification or generalized micromodulus designs and introduce additional parameters \cite{Masoumi2022_MBBPD}. To overcome the constitutive limitations of bond-based models, Silling et al. \cite{Silling2007_StateBased} proposed the SBPD model, and further developed the non-ordinary state-based peridynamic (NOSBPD) model, in which a non-local point-associated deformation gradient is defined so that constitutive relations from the CCM model can be invoked directly \cite{Silling2007_StateBased,Warren2009_NOSBPD}. However, the NOSBPD model generally suffers from zero-energy modes, which may cause numerical instability and errors in the displacement and strain fields. Silling \cite{Silling2017_StabilityCorr}, Li et al. \cite{Li2018_StabilizedNOSBPD}, and Chowdhury et al. \cite{Chowdhury2019_RemoveZeroEnergy} have proposed various stabilization schemes from the perspectives of energy-functional modification, additional force-states, and reformulated constitutive correspondence principle, but these approaches typically introduce extra non-physical stabilizing terms or tuning parameters. Chen \cite{CHEN_2018} argued that the instability of the NOSBPD model arises from the point-associated deformation gradient, which is a neighborhood-averaged approximation and cannot uniquely represent individual bond deformations. Against this background, Chen \cite{CHEN_2018} proposed the bond-associated peridynamic (BAPD) model. In BAPD, a bond-associated deformation gradient is introduced for each bond, so the bond deformation is evaluated directly rather than inferred from a point-associated gradient; this can suppress spurious modes commonly observed in conventional NOSBPD models. Subsequent studies by Chen et al. \cite{Chen2019_BACorrespondence, CHEN_2022} and Madenci et al. \cite{MADENCI_BA_2019} have systematically examined its stability and convergence, derived the weak formulation, and demonstrated that the bond-associated strategy offers advantages in avoiding zero-energy modes while retaining the ability to directly employ classical continuum constitutive models. Compared with NOSBPD, which computes a single point-associated deformation gradient for each neighborhood and uses it for all bonds, BAPD must compute a corresponding bond-associated deformation gradient for each bond, resulting in a significantly higher computational cost.
	
	An effective approach to improve computational efficiency is the PD-CCM coupled model: The PD model is used only in crack domains, while the remaining domain is modeled by the CCM model. Kilic and Madenci \cite{Kilic2010_PDFEM} proposed an overlap domain method, in which the PD and finite element method (FEM) equations are solved simultaneously in an overlap domain, and displacement compatibility and internal force equilibrium are enforced to achieve a smooth transition between the two models. Liu and Hong \cite{Liu2012_PDFEM} developed a coupling algorithm between the PD points and the FEM, in which transition elements or constraint relations are introduced along the interface to connect peridynamic material points with FE nodes, so that peridynamics is used to describe cracks in local domains while the remaining part of the structure is still modeled by FEM. Bobaru and Ha \cite{Bobaru2011_PDAdaptive} employed local adaptive refinement in combination with a variable horizon strategy, using a finer discretization and larger horizons near cracks and a coarser discretization and smaller horizons away from cracks, thereby controlling the overall number of degrees of freedom while preserving the automatic crack propagation capability. Zaccariotto et al. \cite{Zaccariotto2017_PDFE} coupled PD points to FE via an enhanced interface algorithm, in which constraints or bridging elements are used at the interface to enforce displacement continuity and force transfer, thereby achieving effective coupling between PD and CCM models with different levels of discretization. Several researchers have argued that an effective PD-CCM coupling should ensure equivalence between the effective stiffness tensor of the PD model and that of the CCM model. Based on the equivalence of strain energy density, Lubineau et al. \cite{Lubineau2012_MorphingNonlocalLocal} constructed a spatially varying ``Morphing" function so that the BBPD model and the CCM model undergo a transition of material properties within a unified formulation. Subsequently, this method was extended to the coupling between the SBPD model and the CCM model \cite{Han2016_MorphingSBPD,Li2024_AdaptiveNOSBPD_CCM}, and ``Morphing" coupling frameworks were developed that can adaptively adjust the PD domains according to the evolution of cracks \cite{Azdoud2014_MorphingAdaptiveFracture,Wang2021_StrengthInducedPD}. Such PD-CCM models in ``Morphing" frameworks have also been implemented in commercial software \cite{HWP2025,Han2025_PERISOFT}. In addition, the PD-CCM model reduces boundary effects, and enables surface tractions to be imposed on the CCM model \cite{DElia2022_LtNReview}.
	
	This study extends the existing ``Morphing"-type PD-CCM coupling framework \cite{GILLERS2012, HAN2021, Li2024_AdaptiveNOSBPD_CCM} to the BAPD model for quasi-static and dynamic fracture analysis. Under the assumptions of homogeneous infinitesimal deformation and linear elasticity, the correction factor of the BAPD force-state is reformulated based on the pointwise equivalence of strain energy density between the BAPD and CCM models. On this basis, a coupled model is developed, in which a ``Morphing" function provides a smooth transition of the stiffness between the two models. The BAPD model is employed in crack domains to describe crack initiation and propagation. Its bond-associated deformation gradient permits the use of constitutive relations from CCM and makes the model less susceptible to the zero-energy modes commonly observed in conventional NOSBPD. The remaining domain is modeled by CCM to reduce the computational cost. Finally, the solution accuracy, fracture simulation capability, and computational efficiency of the proposed model are evaluated through quasi-static and dynamic examples involving two- and three-dimensional problems.
	
	The remainder of this paper is organized as follows. Section \ref{S2} reviews the fundamental theories of the CCM model and the PD model, and compares the constructions of the BAPD and NOSBPD models. Section \ref{imporve_BA} proposes a new correction factor to improve the BAPD model. Section \ref{S4} develops a coupled model, referred to as the BAPD-CCM model, under a ``Morphing" framework based on strain energy equivalence. Section \ref{S5} presents explicit algorithms for quasi-static and dynamic problems, and introduces two adaptive algorithms for expanding the PD domain. Section \ref{S6} evaluates the accuracy of the improved BAPD model and the BAPD-CCM model. Section \ref{S7} demonstrates the capability of the coupling model to simulate fracture problems through several numerical examples. Finally, Section \ref{S8} summarizes the main contributions.
	
	\section{Review of the CCM and PD models}\label{S2}
	This section reviews the CCM and PD models. The conventional NOSBPD model suffers from zero-energy modes, which several researchers attribute to deficiencies in the construction of the non-local deformation gradient \cite{CHEN_2018, GUXIN_2019, TIAN_2022}. To address this issue, BAPD introduces a bond-associated deformation gradient that effectively eliminates the zero-energy modes.
	\subsection{The CCM theory}
	In the Lagrangian framework (material description), the equation of motion within the CCM model is derived from the principle of linear momentum conservation, establishing a relation between the acceleration of a material point and the stresses and body forces acting upon it. For a material point $\bm{x}$ in the reference configuration $\Omega_{\mathrm{R}}$, the governing equation is expressed as:
	\begin{equation}\label{CCM_motion}
		\rho(\bm{x})\ddot{\bm{u}}(\bm{x},t) = \nabla \cdot \bm{P}^{\mathrm{C}}(\bm{x},t) + \bm{b}(\bm{x},t), \ \forall \bm{x} \in \Omega_{\mathrm{R}}, \ \forall t > 0,
	\end{equation}
	where $\rho$ is the material density, $\bm{u}$ denotes the displacement vector, and $\ddot{(\,\cdot\,)}$ represents the second derivative with respect to time, such that $\ddot{\bm{u}}(\bm{x},t)$ is the acceleration of point $\bm{x}$ at time $t$. The term $\bm{b}$ represents the body force acting on point $\bm{x}$. $\nabla\!\cdot\!\bm{P}^{\mathrm{C}}$ denotes the divergence of the first Piola-Kirchhoff stress tensor $\bm{P}^{\mathrm{C}}$, which is work conjugate to the deformation gradient $\bm{F}^{\mathrm{C}}$ in the CCM model:
	\begin{equation}\label{u_grad}
		\bm{F}^{\mathrm{C}}(\bm{x},t) = \bm{I} + \bm{L}^{\mathrm{C}}(\bm{x}, t),
	\end{equation}
	$\bm{I}$ denotes the identity tensor. $\bm{L}^{\mathrm{C}}(\bm{x}, t) = \nabla\bm{u}(\bm{x}, t)$ is the displacement gradient tensor in the CCM model and $\bm{u}$ is the displacement vector.
	\subsection{The PD theory}
	At time $t$, the equation of motion of the PD model in the reference configuration $\Omega_{\mathrm{R}}$ at point $\bm{x}$ can be expressed as:
	\begin{equation}\label{PD_motion}
		\rho(\bm{x})\ddot{\bm{u}}(\bm{x},t) = \int_{H_{\bm{x}}} \{ {\underline{\bm{T}}[\bm{x},t]\langle\bm{\xi}\rangle - \underline{\bm{T}}[\bm{x}',t]\langle-\bm{\xi}\rangle} \} \mathrm{d}V_{\bm{\xi}}+\bm{b}(\bm{x},t), \ \forall \bm{x} \in \Omega_{\mathrm{R}}, \ \forall t > 0,
	\end{equation}
	where $H_{\bm{x}}$ denotes the neighborhood of point $\bm{x}$, whose radius $\delta$ is referred to as the horizon. The vector $\bm{\xi}$ is termed the bond vector, defined as $\bm{\xi} = \bm{x}' - \bm{x}$. $\underline{\bm{T}}$ represents the force-state, whose definition varies among different PD models.
	\subsubsection{The NOSBPD model}
	For the conventional NOSBPD model, the force-state is defined as:
	\begin{equation}\label{T}
		\underline{\bm{T}}[\bm{x},t]\langle\bm{\xi}\rangle = \omega\langle|\bm{\xi}|\rangle\bm{P}^{\mathrm{P}}(\bm{x},t) \cdot \bm{K}^{-1}(\bm{x}) \cdot \underline{\bm{X}}\langle\bm{\xi}\rangle,
	\end{equation}
	where the function $\omega$ denotes the influence function, a non-negative function of the bond that depends only on the bond length $|\bm{\xi}|$. $\bm{K}$ represents the shape tensor of point $\bm{x}$ and is defined as:
	\begin{equation}\label{K}
		\bm{K}(\bm{x}) = \int_{H_{\bm{x}}}\omega\langle|\bm{\xi}|\rangle\underline{\bm{X}}\langle\bm{\xi}\rangle \otimes \underline{\bm{X}}\langle\bm{\xi}\rangle \mathrm{d}V_{\bm{\xi}}.
	\end{equation}
	where $\bm{P}^{\mathrm{P}}$ denotes the first Piola--Kirchhoff stress, and $\bm{F}^{\mathrm{P}}$ denotes the work-conjugate non-local deformation gradient, which is defined in the NOSBPD model to characterize the deformation at a material point. To distinguish it from the bond-associated deformation gradient introduced later, it is referred to here as the point-associated deformation gradient. The point-associated deformation gradient $\bm{F}^{\mathrm{P}}$ in NOSBPD model is defined as:
	\begin{equation}\label{F}
		\bm{F}^{\mathrm{P}}(\bm{x},t) = \Big(\int_{H_{\bm{x}}}\omega\langle|\bm{\xi}|\rangle\underline{\bm{Y}}\langle\bm{\xi}\rangle\otimes\underline{\bm{X}}\langle\bm{\xi}\rangle \mathrm{d} V_{\bm{\xi}}\Big)\cdot \bm{K}^{-1}(\bm{x}),
	\end{equation}
	where $\underline{\bm{X}}$ represents the relative position vector state between two points in the reference configuration $\Omega_{\mathrm{R}}$ and is defined as:
	\begin{equation}\label{X}
		\underline{\bm{X}}\langle\bm{\xi}\rangle = \bm{\xi} = \bm{x'} - \bm{x},
	\end{equation}
	here, the angle brackets $\langle\bm{\xi}\rangle$ denote the state associated with the bond $\bm{\xi}$. The relative position vector state $\underline{\bm{Y}}$ in the current configuration $\Omega_{\mathrm{C}}$ is defined as:
	\begin{equation}\label{Y}
		\underline{\bm{Y}}\langle\bm{\xi}\rangle = \bm{\xi} + \bm{\eta},
	\end{equation}
	 where $\bm{\eta}$ is the relative displacement vector, and the corresponding state $\underline{\bm{U}}$ is defined as:
	\begin{equation}\label{U}
		\underline{\bm{U}}\langle\bm{\xi}\rangle = \bm{\eta} = \bm{u}(\bm{x'},t)-\bm{u}(\bm{x},t).
	\end{equation}

	Due to the weighted-averaging nature, $\bm{F}^{\mathrm{P}}$ partially neglects the local coherence, orientation, and peak characteristics of deformation \cite{GUXIN_2019}. In general, $\bm{F}^{\mathrm{P}}(\bm{x},t) \ne \bm{F}^{\mathrm{P}}(\bm{x'},t)$, thus the point-associated deformation gradient tensor generally does not necessarily satisfy the kinematic constraint \cite{GUXIN_2019}:
	\begin{equation}\label{F_kc}
		\bm{F}^{\mathrm{P}}(\bm{x},t) \cdot \bm{\xi} \ne \bm{F}^{\mathrm{P}}(\bm{x'},t) \cdot \bm{\xi}.
	\end{equation}
	This means that the deformation of the identical bond, when predicted from the point-associated deformation gradients at $\bm{x}$ and $\bm{x}'$, is generally inconsistent.
	
	\subsubsection{The BAPD model}

	To address the limitations of the point-associated deformation gradient in the conventional NOSBPD model, Chen \cite{CHEN_2018} proposed a model that associates the deformation gradient with individual bonds, referred to as the BAPD model. The force-state in the BAPD model is defined as:
	\begin{figure}[!htbp]
		\centering
		\includegraphics[width=0.8\textwidth]{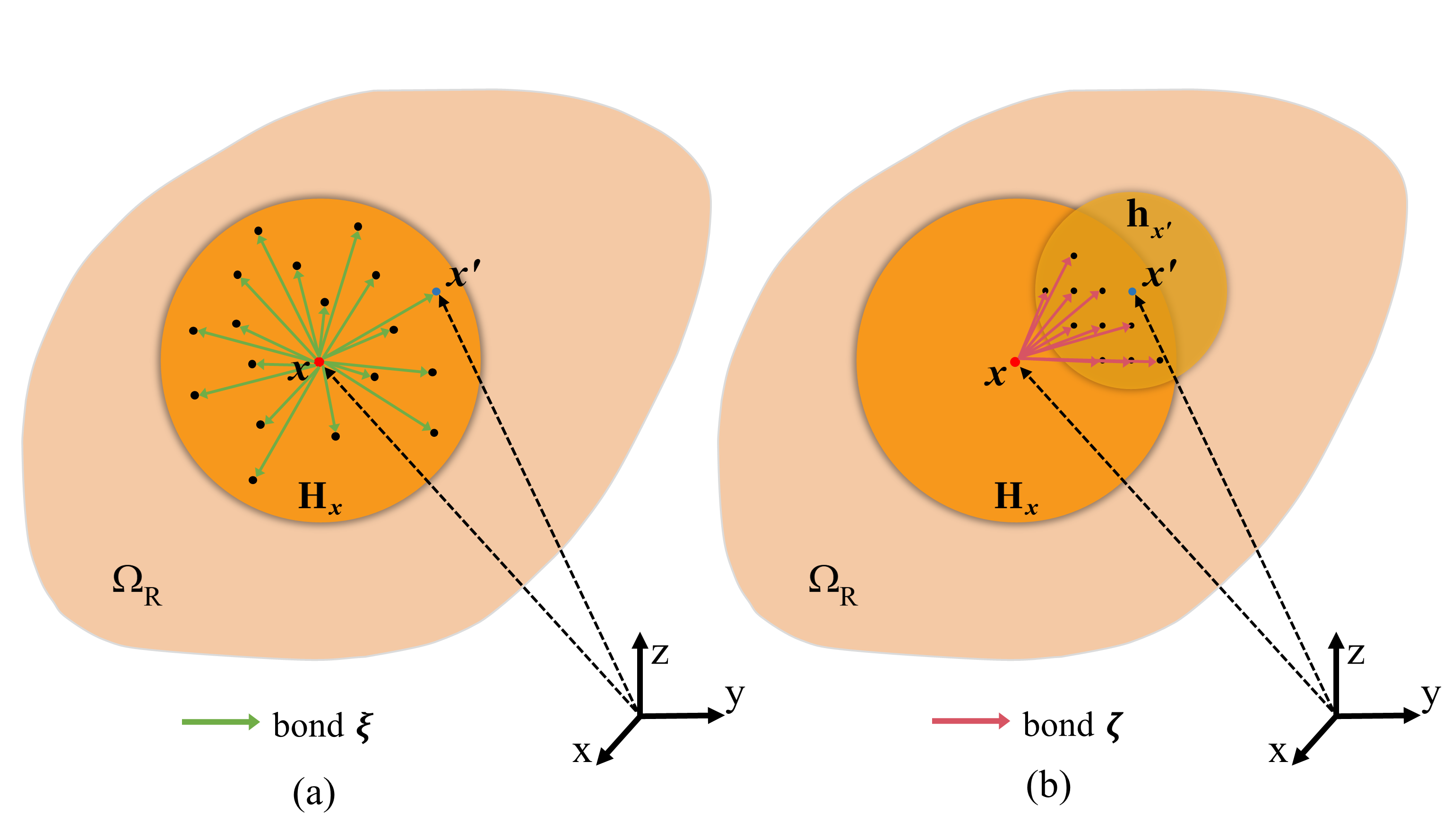}
		\caption{Comparison of the deformation gradient configurations of NOSBPD (a) and BAPD (b) models.}
		\label{BA PD model}
	\end{figure}
	\begin{equation}\label{T_BA}
		\underline{\bm{T}}[\bm{x},t]\langle\bm{\xi}\rangle = \chi_{\bm{\xi}}\underline{\bm{T}}_{\bm{\xi}}[\bm{x},t]\langle\bm{\xi}\rangle, 
	\end{equation}
	where $\underline{\bm{T}}_{\bm{\xi}}$ denotes the bond-associated force-state. The force-state in the BAPD model is obtained by applying a correction to the bond-associated force-state. The corresponding correction factor, $\chi_{\bm{\xi}}^v\in[0,1]$, is commonly defined as the ratio of the bond-associated family volume to the neighborhood volume \cite{CHEN_2018}:
	\begin{equation}\label{chi_xi}
		\chi_{\bm{\xi}} = \chi_{\bm{\xi}}^v = \dfrac{\int_{H_{\bm{x}}\cap h_{\bm{x'}}} 1\,\mathrm{d}V_{\bm{\xi}}}{\int_{H_{\bm{x}}}1\,\mathrm{d}V_{\bm{\xi}}}.
	\end{equation}
     The bond-associated force-state $\underline{\bm{T}}_{\bm{\xi}}$ can be defined as:
	\begin{equation}\label{T_BA_1}
		\underline{\bm{T}}_{\bm{\xi}}[\bm{x},t]\langle\bm{\xi}\rangle = \omega\langle|\bm{\xi}|\rangle \bm{P}_{\bm{\xi}} \cdot \bm{K_{\xi}}^{-1} \cdot \underline{\bm{X}}\langle\bm{\xi}\rangle,
	\end{equation}
	where $\bm{K}_{\bm{\xi}}$ denotes the bond-associated shape tensor, which is defined as:
	\begin{equation}\label{K_xi}
		\bm{K}_{\bm{\xi}} =  \int_{H_{\bm{x}} \cap h_{\bm{x'}}}\omega\langle|\bm{\zeta}|\rangle \underline{\bm{X}}\langle\bm{\zeta}\rangle \otimes \underline{\bm{X}}\langle\bm{\zeta}\rangle  \mathrm{d}V_{\bm{\zeta}},
	\end{equation}
	where $\bm{P}_{\bm{\xi}}$ denotes the bond-associated first Piola-Kirchhoff stress tensor of bond $\bm{\xi}$ and the work-conjugate bond-associated deformation gradient $ \bm{F}_{\bm{\xi}} $ is defined as (Fig. \ref{BA PD model} (b)):
	\begin{equation}\label{F_xi}
		\bm{F_{\xi}} = \Big(\int_{H_{\bm{x}} \cap h_{\bm{x'}}}\omega\langle|\bm{\zeta}|\rangle\underline{\bm{Y}}\langle\bm{\zeta}\rangle \otimes \underline{\bm{X}}\langle\bm{\zeta}\rangle \mathrm{d}V_{\bm{\zeta}}\Big) \cdot \bm{K_{\xi}}^{-1}.
	\end{equation}
    Here, $h_{\bm{x'}}$ represents the bond-associated neighborhood of point $\bm{x'}$. $\bm{\zeta} = \bm{x''} - \bm{x}$, where point $\bm{x''}$ lies within the intersection domain $H_{\bm{x}} \cap h_{\bm{x'}}$. We can define the relationship between $\bm{P}_{\bm{\xi}}$ and $\bm{F}_{\bm{\xi}}$ based on the strain energy density $W_{\bm{\xi}}$ within the domain $H_{\bm{x}} \cap h_{\bm{x'}}$:
	\begin{equation}\label{P_xi}
		\bm{P}_{\bm{\xi}} \equiv \frac{\partial W_{\bm{\xi}}}{\partial \bm{F}_{\bm{\xi}}}.
	\end{equation} 

	The point-associated deformation gradient $\bm{F}^{\mathrm{P}}$ is constructed through an averaging operation over the neighborhood, so that multiple bonds within a family share the same deformation gradient. As a result, deformation perturbations of individual bonds may be weakened or even smeared out during the averaging process, which reduces the model's ability to distinguish single-bond deformation and may lead to zero-energy modes. In contrast, the bond-associated deformation gradient $\bm{F}_{\bm{\xi}}$ is defined in a one-to-one manner for each bond. Therefore, even if a deformation perturbation occurs in a single bond, it can still be reflected in the model. Compared with the NOSBPD model, $\bm{F}_{\bm{\xi}}$ improves the model's ability to resolve single bond deformation, thereby alleviating or even avoiding zero-energy modes.
	\subsection{Simplification of the CCM and BAPD models}

	In this subsection, the CCM and BAPD models are linearized under the infinitesimal deformation assumption.
	\subsubsection{Linearization of the CCM model}

	Under the infinitesimal deformation assumption, the first Piola-Kirchhoff stress tensor $\bm{P}^{\mathrm{C}}$ in CCM can be approximated by the Cauchy stress tensor $\bm{\sigma}^{\mathrm{C}}$ as:
	\begin{equation}\label{P_sigma}
		\bm{P}^{\mathrm{C}}(\bm{x}, t) \approx \bm{\sigma}^{\mathrm{C}}(\bm{x},t),
	\end{equation}
	under the generalized Hooke's law, the Cauchy stress tensor satisfies:
	\begin{equation}\label{Hooke}
		\bm{\sigma}^{\mathrm{C}}(\bm{x},t) = \mathbb{E}^{0} : \bm{\varepsilon}^{\mathrm{C}}(\bm{x},t),
	\end{equation}
	here, $\mathbb{E}^{0}$ denotes the elastic stiffness tensor, and $\bm{\varepsilon}^{\mathrm{C}}$ is the infinitesimal strain tensor, which is defined as:
	\begin{equation}\label{Small_strain}
		\bm{\varepsilon}^{\mathrm{C}}(\bm{x},t) = \mathrm{sym}(\bm{L}^{\mathrm{C}}(\bm{x}, t)) = \frac{1}{2}[\bm{L}^{\mathrm{C}}(\bm{x}, t) + {}^{T}\bm{L}^{\mathrm{C}}(\bm{x}, t)] ,
	\end{equation}
	where $\mathrm{sym}(\cdot)$ denotes the symmetrization operator. 
	
	To sum up, under the assumption of infinitesimal deformation, Eq.~\eqref{CCM_motion} can then be written as:
	\begin{equation}\label{linear_CCM_motion}
		\rho(\bm{x})\ddot{\bm{u}}(\bm{x},t) = \nabla \cdot \bm{\sigma}^{\mathrm{C}}(\bm{x},t) + \bm{b}(\bm{x},t), \forall \bm{x} \in \Omega_{\mathrm{R}}, \forall t > 0.
	\end{equation}
	\subsubsection{Linearization of the BAPD model}

	In the BAPD model, the relationship between the bond-associated first Piola-Kirchhoff stress tensor $\bm{P}_{\bm{\xi}}$ and the bond-associated deformation gradient $\bm{F}_{\bm{\xi}}$ is formulated by analogy with that in CCM. Accordingly, following the linearization of the CCM formulation, $\bm{P}_{\bm{\xi}}$ under the infinitesimal deformation assumption can be written as:
	\begin{equation}\label{P_xi_sigma}
		\bm{P}_{\bm{\xi}} \approx \bm{\sigma}_{\bm{\xi}},
	\end{equation}
	where $\bm{\sigma}_{\bm{\xi}}$ denotes the bond-associated Cauchy stress tensor. The relationship between $\bm{\sigma}_{\bm{\xi}}$ and the bond-associated infinitesimal strain $\bm{\varepsilon}_{\bm{\xi}}$ is given as follows:
	\begin{equation}\label{Hooke_xi}
		\bm{\sigma}_{\bm{\xi}} = \mathbb{E}^{0} : \bm{\varepsilon}_{\bm{\xi}},
	\end{equation}
	here, $\bm{\varepsilon}_{\bm{\xi}}$ is defined as:
	\begin{equation}\label{small_strain_xi}
		\bm{\varepsilon}_{\bm{\xi}} = \mathrm{sym}(\bm{L}_{\bm{\xi}}),
	\end{equation}
	Similar to Eq.~\eqref{u_grad}, the bond-associated displacement gradient $\bm{L}_{\bm{\xi}}$ can also be obtained from Eq.~\eqref{F_xi} as:
	\begin{equation}\label{U_grad_xi}
		\bm{L}_{\bm{\xi}} = \Big(\int_{H_{\bm{x}} \cap h_{\bm{x'}}} \omega\langle|\bm{\zeta}|\rangle\underline{\bm{U}}\langle\bm{\zeta}\rangle \otimes \underline{\bm{X}}\langle\bm{\zeta}\rangle \mathrm{d}V_{\bm{\zeta}} \Big)\cdot\bm{K}_{\bm{\xi}}^{-1}.
	\end{equation}
	\subsubsection{The kinematically constrained BAPD model}

	In the BAPD model, the size of bond-associated neighborhood $h_{\bm{x'}}$ is important \cite{CHEN_2018}, as it significantly influences the accuracy of the results. Gu et al. \cite{GUXIN_2019} and Tian et al. \cite{TIAN_2022} suggested that the bond-associated neighborhood $h_{\bm{x'}}$ should be isometric to the neighborhood $H_{\bm{x}}$. Such a configuration ensures that, when computing $\bm{\xi}$ and $\bm{-\xi}$, the overlapping domain contains the same set of material points, $H_{\bm{x}} \cap h_{\bm{x'}} = H_{\bm{x'}} \cap h_{\bm{x}}$, $\bm{K}_{\bm{\xi}}$ and $\bm{F}_{\bm{\xi}}$ satisfy:
	\begin{equation}\label{F_K_xi_sym}
		\bm{K}_{\bm{\xi}} = \bm{K}_{-\bm{\xi}}, \ \ \bm{F}_{\bm{\xi}} = \bm{F}_{-\bm{\xi}},
	\end{equation}
	where the integration domain for $\bm{K}_{-\bm{\xi}}$ and $\bm{F}_{-\bm{\xi}}$ is replaced by $H_{\bm{x'}} \cap h_{\bm{x}}$. The bond $\bm{\xi}$ satisfies:
	\begin{equation}\label{F_xi_kc}
		\bm{\xi} + \bm{\eta} = \bm{F}_{\bm{\xi}} \cdot \bm{\xi} = \bm{F}_{-\bm{\xi}} \cdot \bm{\xi}.
	\end{equation}
	Thus the kinematically constrained BAPD model avoids the issue in NOSBPD model like Eq. (\ref{F_kc}). The bond-associated Cauchy stress tensor satisfies:
	\begin{equation}\label{sigma_xi_sym}
		\bm{\sigma}_{\bm{\xi}} = \bm{\sigma}_{-\bm{\xi}}.
	\end{equation}
	Finally, we obtain:
	\begin{equation}\label{T_xi_sym}
		\begin{aligned}
			\underline{\bm{T}}[\bm{x},t]\langle\bm{\xi}\rangle & = \chi_{\bm{\xi}}  \omega\langle|\bm{\xi}|\rangle \bm{\sigma}_{\bm{\xi}} \cdot \bm{K_{\xi}}^{-1}(\bm{x}) \cdot \underline{\bm{X}}\langle\bm{\xi}\rangle\\
			& = \chi_{-\bm{\xi}} \omega\langle|-\bm{\xi}|\rangle \bm{\sigma}_{-\bm{\xi}} \cdot \bm{K_{-\xi}}^{-1}(\bm{x}) \cdot \underline{\bm{X}}\langle\bm{\xi}\rangle = -\underline{\bm{T}}[\bm{x'},t]\langle-\bm{\xi}\rangle.
		\end{aligned}
	\end{equation}
	\section{An improved BAPD model}\label{imporve_BA}
	For a linearly elastic material undergoing homogeneous infinitesimal deformation, this study reformulates the correction factor in the BAPD force state by using the equivalence of strain energy density between BAPD and CCM at a point. This treatment of establishing model equivalence under homogeneous deformation follows the correspondence definition introduced by Silling et al. \cite{Silling2007_StateBased} for relating NOSBPD to CCM models. It is not intended to limit the application of the model to homogeneous deformation fields.

	Following previous research \cite{SILLING_2000, SILLING_2005, MADENCI_2014, BOBARU_2016}, the strain energy density of the BAPD model is derived by integrating the bond energy density over its neighborhood. Under the above assumption, this strain energy density is equal to its CCM counterpart. In the absence of energy dissipation, the energy density of a single bond $\bm{\xi}$ is defined as:
	\begin{equation}\label{psi_def}
		\psi_{\mathrm{BA}}(\bm{\xi},t)=\int_{\hat{\bm{U}}\langle\bm{\xi}\rangle=\bm{0}}^{\hat{\bm{U}}\langle\bm{\xi}\rangle=\underline{\bm{U}}\langle\bm{\xi}\rangle}
		\big\{\underline{\bm{T}}[\bm{x},t]\langle\bm{\xi}\rangle-\underline{\bm{T}}[\bm{x}',t]\langle-\bm{\xi}\rangle\big\}\cdot \mathrm{d}\hat{\bm{U}}\langle\bm{\xi}\rangle.
	\end{equation}
    The strain energy density at point $\bm{x}$ is defined as
    \begin{equation}\label{W_BA_density_clean}
    	W_{\mathrm{BA}}(\bm{x},t)=\frac12\int_{H_{\bm{x}}}\psi_{\mathrm{BA}}(\bm{\xi},t) \mathrm{d}V_{\bm{\xi}}.
    \end{equation}
	For all points within the neighborhood $H_{\bm{x}}$ under the assumption of uniform deformation, the following approximation is adopted:
	\begin{equation}\label{slow_grad}
		\bm{L}^{\mathrm{C}}(\bm{x}',t) = \bm{L}^{\mathrm{C}}(\bm{x},t) = \bar{\bm{L}},\qquad
		\underline{\bm{U}}\langle\bm{\xi}\rangle=\bar{\bm{L}}\cdot\underline{\bm{X}}\langle\bm{\xi}\rangle,\quad \forall\,\bm{x}'\in H_{\bm{x}}.
	\end{equation}
	Substituting Eq.~\eqref{slow_grad} into Eq.~\eqref{U_grad_xi} gives:
	\begin{equation}\label{L_xi_equals_Lbar}
		\bm{L}_{\bm{\xi}}=\bar{\bm{L}}\cdot\!\left(\int_{H_{\bm{x}}\cap h_{\bm{x'}}}\!\omega\langle|\bm{\zeta}|\rangle\,\underline{\bm{X}}\langle\bm{\zeta}\rangle\otimes\underline{\bm{X}}\langle\bm{\zeta}\rangle \mathrm{d}V_{\bm{\zeta}}\right)\!\cdot\bm{K}_{\bm{\xi}}^{-1}
		=\bar{\bm{L}}.
	\end{equation}
	Thus, we can obtain $\bm{F}_{\bm{\xi}}=\bar{\bm{F}}$. Accordingly: 
	\begin{equation}\label{e_D_e_L}
		\bm{\varepsilon}_{\bm{\xi}}=\mathrm{sym}(\bm{L}_{\bm{\xi}})=\bar{\bm{\varepsilon}}=\mathrm{sym}(\bar{\bm{L}}), \ \  \bm{\sigma}_{\bm{\xi}}=\bar{\bm{\sigma}}.
	\end{equation}
	Hence, the bond-associated strain tensor coincides with the uniform infinitesimal strain $\bm{\bar{\varepsilon}}$. Since the rotational part does not contribute to the energy, $\bar{\bm{L}}$ can be replaced by $\bar{\bm{\varepsilon}}$. Under the linearized assumption, substituting Eq. \eqref{T_BA}, Eq. \eqref{T_xi_sym} and Eq. \eqref{slow_grad} into Eq. \eqref{psi_def} gives:
	\begin{equation}\label{psi_closed_1}
		\begin{aligned}
			\psi_{\mathrm{BA}}(\bm{\xi},t)
			&=
			\int_{\hat{\bm{U}}\langle\bm{\xi}\rangle=\bm{0}}^{\hat{\bm{U}}\langle\bm{\xi}\rangle=\underline{\bm{U}}\langle\bm{\xi}\rangle}\big\{\underline{\bm{T}}[\bm{x},t]\langle\bm{\xi}\rangle-\underline{\bm{T}}[\bm{x}',t]\langle-\bm{\xi}\rangle\big\}\cdot \mathrm{d}\hat{\bm{U}}\langle\bm{\xi}\rangle\\
			&=
			\int_{\hat{\bm{U}}\langle\bm{\xi}\rangle=\bm{0}}^{\hat{\bm{U}}\langle\bm{\xi}\rangle=\underline{\bm{U}}\langle\bm{\xi}\rangle}\big\{2\chi_{\bm{\xi}}  \omega\langle|\bm{\xi}|\rangle \bar{\bm{\sigma}} \cdot \bm{K_{\xi}}^{-1}(\bm{x}) \cdot \underline{\bm{X}}\langle\bm{\xi}\rangle\big\}\cdot \mathrm{d}\hat{\bm{U}}\langle\bm{\xi}\rangle\\
			&=
			\Big[
			\omega\langle|\bm{\xi}|\rangle \chi_{\bm{\xi}}
			\big(\mathbb{E}^{0}:\bar{\bm{\varepsilon}}\big)
			\cdot \bm{K}_{\bm{\xi}}^{-1}
			\cdot
			\big(
			\underline{\bm{X}}\langle\bm{\xi}\rangle
			\otimes
			\underline{\bm{X}}\langle\bm{\xi}\rangle
			\big)
			\Big]
			:
			\bar{\bm{\varepsilon}} \\
			&=
			\bar{\bm{\varepsilon}}
			:
			\Big[
			\mathbb{E}^{0}
			\cdot \bm{K}_{\bm{\xi}}^{-1}
			\cdot
			\Big(
			\omega\langle|\bm{\xi}|\rangle \chi_{\bm{\xi}}
			\underline{\bm{X}}\langle\bm{\xi}\rangle
			\otimes
			\underline{\bm{X}}\langle\bm{\xi}\rangle
			\Big)
			\Big]
			:
			\bar{\bm{\varepsilon}} .
		\end{aligned}
	\end{equation}
	Substituting Eq. \eqref{psi_closed_1} into Eq. \eqref{W_BA_density_clean} yields the final form:
	\begin{equation}\label{W_BA_final_clean}
		W_{\mathrm{BA}}(\bm{x},t)
		=\frac12\,\bar{\bm{\varepsilon}}:\!\left[\int_{H_{\bm{x}}}\Big[\mathbb{E}^{0}\cdot \bm{K}_{\bm{\xi}}^{-1}
		\cdot\big(\omega\langle|\bm{\xi}|\rangle\chi_{\bm{\xi}}  \underline{\bm{X}}\langle\bm{\xi}\rangle\otimes\underline{\bm{X}}\langle\bm{\xi}\rangle\big)\ \Big] \mathrm{d}V_{\bm{\xi}}\right]\!:\bar{\bm{\varepsilon}}.
	\end{equation}
	
	For any uniform infinitesimal strain $\bar{\bm{\varepsilon}}$, the energy equivalence $W_{\mathrm{BA}}(\bm{x},t)=\frac{1}{2}\,\bar{\bm{\varepsilon}}:\mathbb{E}^{0}:\bar{\bm{\varepsilon}}$ holds if and only if:
	\begin{equation}\label{CCM_E0_BA_E0}
		\int_{H_{\bm{x}}}\Big[\mathbb{E}^{0}\cdot \bm{K}_{\bm{\xi}}^{-1}
		\cdot\big(\omega\langle|\bm{\xi}|\rangle\chi_{\bm{\xi}}  \underline{\bm{X}}\langle\bm{\xi}\rangle\otimes\underline{\bm{X}}\langle\bm{\xi}\rangle\big)\ \Big] \mathrm{d}V_{\bm{\xi}} = \mathbb{E}^{0}.
	\end{equation}
	Since $\mathbb{E}^{0}$ is a fourth-order tensor, Eq.~\eqref{CCM_E0_BA_E0} implies that:
	\begin{equation}\label{CCM_E0_BA_E0_old}
		\int_{H_{\bm{x}}}\Big[\bm{K}_{\bm{\xi}}^{-1}
		\cdot\big(\omega\langle|\bm{\xi}|\rangle\chi_{\bm{\xi}}\underline{\bm{X}}\langle\bm{\xi}\rangle\otimes\underline{\bm{X}}\langle\bm{\xi}\rangle\big)\Big] \mathrm{d}V_{\bm{\xi}}=\bm{I}.
	\end{equation}
	Under the assumption of a spherical neighborhood $H_{\bm x}$ with horizon $\delta$ and a radial influence function $\omega=\omega(|\bm\xi|)$, by requiring the bond-associated shape tensor $\bm K_{\bm \xi}$ and the shape tensor $\bm K(\bm x)$ to have the same quadratic form value along an arbitrary bond direction, the following correction factor satisfying Eq.~\eqref{CCM_E0_BA_E0_old} can be derived:
	\begin{equation}\label{chi_a}
		\chi_{\bm{\xi}} = \chi_{\bm{\xi}}^a
		= \frac{\underline{\bm{X}}\langle\bm{\xi}\rangle^{T}\,\bm{K}(\bm{x})^{-1}\,\underline{\bm{X}}\langle\bm{\xi}\rangle}
		{\underline{\bm{X}}\langle\bm{\xi}\rangle^{T}\,\bm{K}_{\bm{\xi}}^{-1}\,\underline{\bm{X}}\langle\bm{\xi}\rangle}.
	\end{equation}
	The detailed proofs are provided in \textbf{Appendix~A}.
	
	\section{Coupling model of BAPD and CCM}\label{S4}
	
	\begin{figure}[!htbp]
		\centering
		\includegraphics[width=0.85\textwidth]{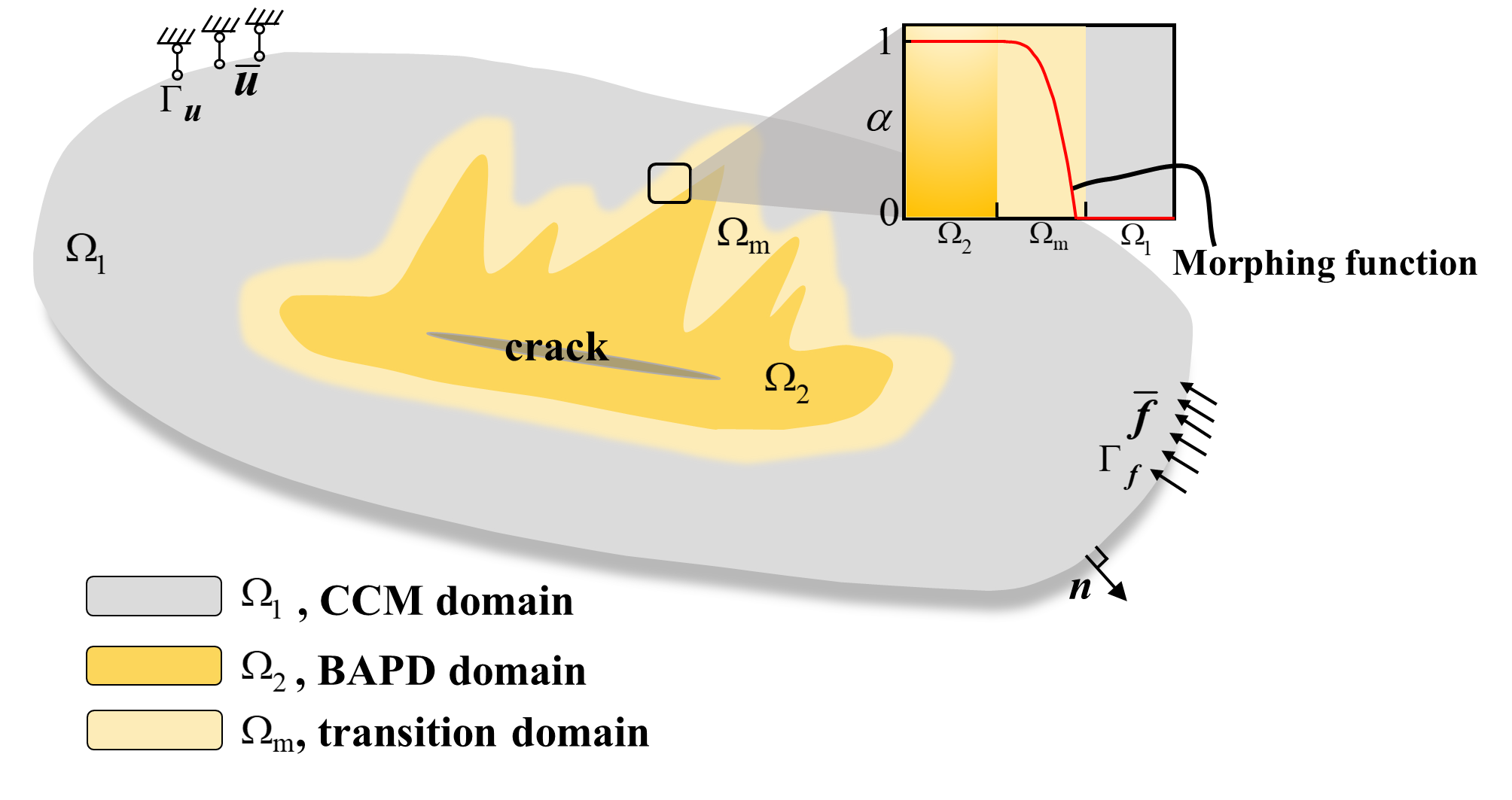}
		\caption{Distribution of the BAPD, CCM, and transition domains in the BAPD-CCM model.}
		\label{coupling model}
	\end{figure}
	This section introduces the BAPD-CCM model, which couples the BAPD and CCM models through strain energy density equivalence. A ``Morphing" strategy is employed to achieve a smooth transition of the stiffness contributions within the transition domain.
	\subsection{Governing equations of the coupling model}
	In the BAPD-CCM model, the computational domain $\Omega$ is partitioned into three domains, $\Omega_{1}$, $\Omega_{2}$, and $\Omega_{m}$. Domain $\Omega_{1}$ is governed by the CCM model, $\Omega_{2}$ by the BAPD model, and $\Omega_{m}$ is the transition domain, as shown in Fig. \ref{coupling model}. It follows that:
	\begin{equation}\label{set_realtion}
		\Omega = \Omega_{1} \cup \Omega_{2} \cup \Omega_{m},	\quad \Omega_{i}\cap\Omega_{j}=\varnothing \ (i\neq j).
	\end{equation}
	Dirichlet displacements $\bar{\bm{u}}$ and external tractions $\overline{\bm{f}}$ are prescribed on	$\Gamma_{\bm{u}}$ and $\Gamma_{\bm{f}}$, respectively. The vector $\bm{n}$ denotes the outward unit normal to the boundary. Under the infinitesimal deformation assumption, the governing equations of the BAPD-CCM model can be written as:\\
	\hspace{-0.65cm}
	$\bullet \quad \textbf{Kinematic admissibility and compatibility:}$
	\begin{equation}\label{CP_GEO_CCM}
		\bm{\varepsilon}^{\mathrm{C}}(\bm{x},t) = \frac12(\bm{L}^{\mathrm{C}}(\bm{x},t)+{}^{T}\bm{L}^{\mathrm{C}}(\bm{x},t)) , \ \forall \bm{x}\in \Omega,  \ \forall t > 0,
	\end{equation}
	\begin{equation}\label{CP_GEO_PD}
		\bm{\eta}=\bm{u}(\bm{x'},t)-\bm{u}(\bm{x},t), \ \forall \bm{x},\bm{x'}\in \Omega,  \ \forall t > 0,
	\end{equation}
	\begin{equation}\label{CP_GEO_BOUN}
		\bm{u}(\bm{x},t)=\bar{\bm{u}}(\bm{x},t), \ \forall \bm{x}\in \Gamma _{\bm{u}},  \ \forall t > 0.
	\end{equation}
	$\bullet \quad \textbf{Dynamic admissibility:}$
	\begin{equation}\label{CP_EQU}
		\rho (\bm{x})\bm{\ddot{u}}(\bm{x},t)=\nabla \cdot \bm{\sigma}^{\mathrm{C}} (\bm{x},t) +\int_{H_{\bm{x}} } \big\{\underline{\bm{T}}[\bm{x},t]\langle\bm{\xi}\rangle-\underline{\bm{T}}[\bm{x}',t]\langle-\bm{\xi}\rangle\big\} \mathrm{d}V_{\bm{\xi}}+\bm{b}(\bm{x},t), \ \forall \bm{x}\in \Omega ,   \ \forall t > 0,
	\end{equation}
	\begin{equation}\label{CP_EQU_BOUN}
		\bm{\sigma}^{\mathrm{C}} (\bm{x},t)\cdot \bm{n}(\bm{x},t) = \overline{\bm{f}}, \ \forall \bm{x}\in \Gamma _{\bm{f}},   \ \forall t > 0.
	\end{equation}
	$\bullet \quad \textbf{Constitutive equations:}$
	\begin{equation}\label{CP_CON_CCM}
		\bm{\sigma}^{\mathrm{C}}(\bm{x},t) = \mathbb{E}^{\mathrm{C}}(\bm{x},t) : \bm{\varepsilon}^{\mathrm{C}}(\bm{x},t), \ \forall \bm{x}\in \Omega,   \ \forall t > 0,
	\end{equation}
	\begin{equation}\label{CP_CON_PD}
		\underline{\bm{T}}[\bm{x},t]\langle\bm{\xi}\rangle = \chi_{\bm{\xi}}\omega\langle|\bm{\xi}|\rangle \mathbb{E}^{\mathrm{P}}(\bm{x},t) : \bm{\varepsilon}_{\bm{\xi}} \cdot \bm{K_{\xi}}^{-1} \cdot \underline{\bm{X}}\langle\bm{\xi}\rangle, \ \forall \bm{x}\in \Omega,   \ \forall t > 0.
	\end{equation}
	where $\mathbb{E}^{\mathrm{C}}(\bm{x},t)$ and $\mathbb{E}^{\mathrm{P}}(\bm{x},t)$ denote the stiffness tensors of the CCM model and the BAPD model, respectively.
	\subsection{Derivation of the stiffness tensor based on strain energy density equivalence}
	To achieve a smooth transition between the stiffness tensors of the two models, a ``Morphing" function, $\alpha(\bm{x},t)\in[0,1]$, is introduced, and:
	\begin{equation}\label{E0_Epd}
		\mathbb{E}^{\mathrm{P}}(\bm{x},t) = \alpha(\bm{x}, t)\mathbb{E}^{0},
	\end{equation}
	while the remaining stiffness contribution is assigned to the CCM model through $\mathbb{E}^{\mathrm{C}}(\bm{x},t)$. Accordingly, $\alpha$ and $\mathbb{E}^{\mathrm{C}}$ together determine the relative contributions of the BAPD and CCM models at each material point.
	
	\noindent\textbf{Case 1:} When $\alpha(\bm{x}, t)=0$, and $\alpha(\bm{x}', t)=0, \forall \bm{x'} \in H_{\bm{x}}$, the point $\bm{x}$ is located in the CCM domain, $\bm{x} \in \Omega_{1}$. The total stiffness is provided by the CCM model. The strain energy density can be expressed as:
	\begin{equation}\label{W_CCM}
		W_{\mathrm{CCM}}(\bm{x},t) = \frac12\bar{\bm{\varepsilon}}(\bm{x},t) : \mathbb{E}^{0}: \bar{\bm{\varepsilon}}(\bm{x},t).
	\end{equation}
	\noindent\textbf{Case 2:} When $\alpha(\bm{x}, t)=1$, and $\alpha(\bm{x}', t)=1, \forall \bm{x'} \in H_{\bm{x}}$, the point $\bm{x}$ is located in the BAPD domain, $\bm{x} \in \Omega_{2}$. The BAPD model contributes the total stiffness, $W_{\mathrm{BA}}(\bm{x},t)$ follows from Eq. \eqref{W_BA_final_clean}.\\
	\noindent\textbf{Case 3:} When $0\le\alpha(\bm{x}, t)\le 1$, and $ 0 < \alpha(\bm{x}', t) < 1, \exists \bm{x'} \in H_{\bm{x}}$, the point $\bm{x}$ is located in the transition domain, $\bm{x} \in \Omega_{m}$. The total stiffness is jointly contributed by the CCM model and the BAPD model:
	\begin{equation}\label{W_tran}
		\begin{aligned}
			&W_{\mathrm{Tran}}(\bm{x},t) = \frac12\bar{\bm{\varepsilon}}(\bm{x},t) : \mathbb{E}^{\mathrm{C}}(\bm{x},t): \bar{\bm{\varepsilon}}(\bm{x},t) 
			\\ &+ \frac12\,\bar{\bm{\varepsilon}}(\bm{x},t):\Big(\int_{H_{\bm{x}}}\frac12(\mathbb{E}^{\mathrm{P}}(\bm{x},t)+\mathbb{E}^{\mathrm{P}}(\bm {x'},t))\cdot \bm{K}_{\bm{\xi}}^{-1}
			\cdot\big(\omega\langle|\bm{\xi}|\rangle\,\chi_{\bm{\xi}}\underline{\bm{X}}\langle\bm{\xi}\rangle\otimes\underline{\bm{X}}\langle\bm{\xi}\rangle\big)\,\mathrm{d}V_{\bm{\xi}}\Big):\bar{\bm{\varepsilon}}(\bm{x},t)
			\\ & = \frac12\bar{\bm{\varepsilon}}(\bm{x},t) : \Big[\mathbb{E}^{\mathrm{C}}(\bm{x},t) + \Big(\int_{H_{\bm{x}}}\frac12(\mathbb{E}^{\mathrm{P}}(\bm{x},t)+\mathbb{E}^{\mathrm{P}}(\bm {x'},t))\cdot \bm{K}_{\bm{\xi}}^{-1}\cdot\big(\omega\langle|\bm{\xi}|\rangle\,\chi_{\bm{\xi}}\underline{\bm{X}}\langle\bm{\xi}\rangle\otimes\underline{\bm{X}}\langle\bm{\xi}\rangle\big)\,\mathrm{d}V_{\bm{\xi}}\Big) \Big]: \bar{\bm{\varepsilon}}(\bm{x},t).
		\end{aligned}
	\end{equation}
	The above equation holds for any $\bar{\bm{\varepsilon}}$. From the equivalence of the strain energy density, that is, from the equivalence between Eqs.~\eqref{W_CCM} and \eqref{W_tran}, it follows that:
	\begin{equation}\label{E_CCM}
		\mathbb{E}^{\mathrm{C}}(\bm{x},t) = 	\mathbb{E}^{0} \cdot \Big[ \bm{I} - \Big(\int_{H_{\bm{x}}} \frac12(\alpha(\bm{x}, t) + \alpha(\bm{x'}, t))\bm{K}_{\bm{\xi}}^{-1}
		\cdot\big(\omega\langle|\bm{\xi}|\rangle\,\chi_{\bm{\xi}}\underline{\bm{X}}\langle\bm{\xi}\rangle\otimes\underline{\bm{X}}\langle\bm{\xi}\rangle\big)\,\mathrm{d}V_{\bm{\xi}}\Big) \Big].
	\end{equation}
    \subsection{Damage description in BAPD-CCM model}

    In this study, a critical-stretch criterion of the bond is adopted. For brittle materials, once the stretch of a bond satisfies this criterion, the bond breaks irreversibly. The state of a bond is characterized by a scalar variable $\mu(\boldsymbol{\xi}, t)$:
    \begin{equation}\label{bond_break}
		\mu(\boldsymbol{\xi},t)=
		\begin{cases}
			0, & s(\boldsymbol{\xi},t)\ge s_c,\\[2pt]
			1, & \text{otherwise},
		\end{cases}
    \end{equation}
     where $s_c$ is the critical bond stretch determined from the material fracture energy $G_{c}$. To link microscopic bond-breaking to macroscopic fracture, a scalar variable called damage, $\phi(\bm{x}, t)$, is introduced for each material point $\bm{x}$, which satisfies $0 \le \phi(\bm{x}, t) \le 1$. However, in the BAPD-CCM model, bond-breaking is confined to the BAPD domain $\Omega_2$, so that the damage $\phi(\bm{x}, t)$ satisfies:
    \begin{equation}\label{dam_point}
	\phi(\bm{x},t)=
	\begin{cases}
		\displaystyle
		\frac{\int_{H_{\bm{x}}}(1-\mu(\boldsymbol{\xi},t))\,\mathrm{d}V_{\bm \xi}}{\int_{H_{\bm{x}}}1\,\mathrm{d}V_{\bm \xi}}, & \bm{x}\in\Omega_{2},\\[10pt]
		0, & \bm{x}\in\Omega\setminus\Omega_{2}.
	\end{cases}
    \end{equation}
	\section{Numerical algorithm}\label{S5}
	This section presents the numerical solution algorithm for the proposed model, including the spatial and temporal discretizations. The CCM model is solved by the FEM, whereas the BAPD model is evaluated by meshfree nodal integration. In terms of time discretization, explicit schemes are adopted for both quasi-static and dynamic problems, with adaptive dynamic relaxation (ADR) used for quasi-static cases and the central difference method for dynamic cases. In addition, strength-induced \cite{Wang2021_StrengthInducedPD, Li2024_AdaptiveNOSBPD_CCM} and bond-breaking-induced \cite{Azdoud2014_MorphingAdaptiveFracture} criteria are introduced to adaptively expand the BAPD domain during the analysis. For brevity, the detailed formulations of time discretization and adaptive algorithms are omitted here, as only the spatial discretization of the coupled model is discussed in this section.
	
	\begin{figure}[!htbp]
		\centering
		\includegraphics[width=0.8\textwidth]{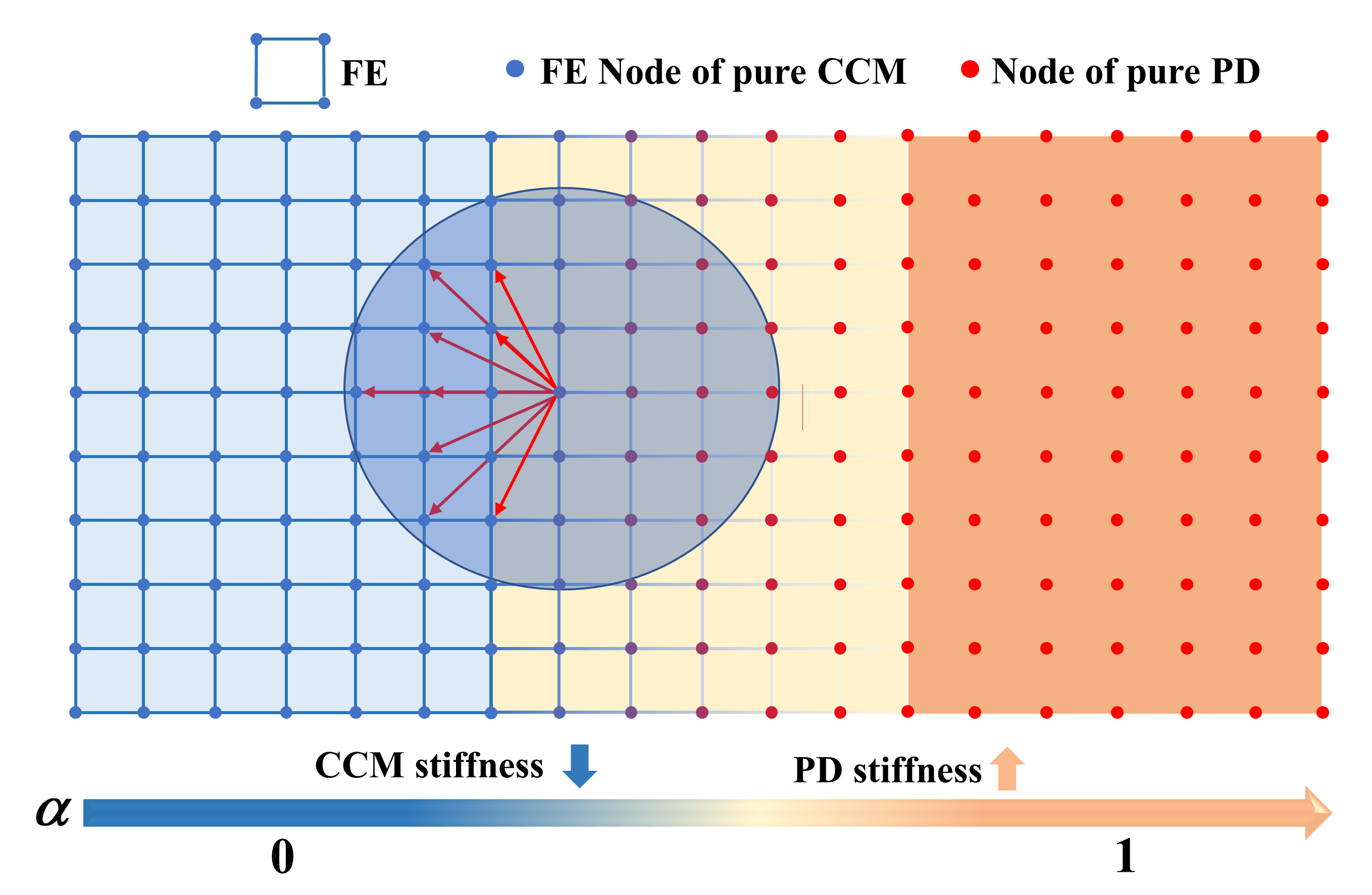}
		\caption{Discretization of the BAPD-CCM model and stiffness transition from CCM to BAPD.}
		\label{CP_model_dis}
	\end{figure}
	
	The computational model is discretized into $m$ elements and $n$ nodes. As illustrated in Fig. \ref{CP_model_dis}, when the node is located in $\Omega_1$, node connectivity is preserved and the internal forces are computed by the CCM model via the FEM. When the node is located in $\Omega_2$, all connectivity associated with node $i$ is removed, the element nodes are converted into independent PD particles, and the internal forces are provided by the BAPD model. When the node is located in $\Omega_m$, both element- and node-wise contributions coexist, and each model contributes the internal force. It is worth noting that, since the element nodes and the PD particles in the transition domain are collocated, displacement continuity across the coupling interface is naturally ensured.

	\section{Model validation}\label{S6}
	This section first evaluates the performance of the improved BAPD model proposed in Section~\ref{imporve_BA}. Then the solution accuracy of the proposed BAPD-CCM model is verified in both quasi-static and dynamic cases. To ensure satisfaction of the kinematic constraint inherent in the adopted BAPD model, the bond-associated neighborhood $h_{\bm{x'}}$ is taken to coincide with the neighborhood $H_{\bm{x}}$ in the subsequent cases.
	\subsection{Evaluation of improved BAPD model}
    \begin{table}[htbp]
    	\centering
    	\caption{Principal value $\lambda$ for different correction factors under different horizons.}
    	\label{tab:lambda_chi}
    	\begin{tabular}{c c c}
    		\toprule
    		$\delta/\Delta x$ 
    		& $\lambda$ for $\chi_v$ 
    		& $\lambda$ for $\chi_a$ \\
    		\midrule
    		3 & 1.090 & 1.000 \\
    		4 & 1.049 & 1.000 \\
    		5 & 1.041 & 1.000 \\
    		\bottomrule
    	\end{tabular}
    \end{table}
	A numerical verification is performed: under various choices of $\chi_{\bm{\xi}}$, the energy equivalence between the BAPD and CCM models is assessed, and compliance with Eq.~\eqref{CCM_E0_BA_E0_old} is quantified by the corresponding error. Since $\chi_{\bm{\xi}}$ is a scalar and $\omega=\omega(|\bm{\xi}|)$ depends only on $|\bm{\xi}|$, the integrand has the radially symmetric structure. Over the sphere $H_{\bm{x}}$, all off-diagonal terms cancel by symmetry and the diagonal entries are identical. Hence:
	\begin{equation}
		\int_{H_{\bm{x}}}
		\bm{K}_{\bm{\xi}}^{-1} \cdot \!\big(\omega\langle|\bm{\xi}|\rangle\,\chi_{\bm{\xi}}\;
		\underline{\bm{X}}\langle\bm{\xi}\rangle\!\otimes\!\underline{\bm{X}}\langle\bm{\xi}\rangle\big)\,\mathrm{d}V_{\bm{\xi}}
		= \lambda \bm{I},
	\end{equation}
	and
	\begin{equation}
		\lambda=\frac{1}{d_{\mathrm{dim}}}\operatorname{tr}\!\left[
		\int_{H_{\bm{x}}}\bm{K}_{\bm{\xi}}^{-1} \cdot \!\big(\omega\langle|\bm{\xi}|\rangle\,\chi_{\bm{\xi}}\;
		\underline{\bm{X}}\langle\bm{\xi}\rangle\!\otimes\!\underline{\bm{X}}\langle\bm{\xi}\rangle\big)\,\mathrm{d}V_{\bm{\xi}}
		\right].
	\end{equation}
    where the $d_{\mathrm{dim}}$ is the spatial dimension. According to Eq.~\eqref{CCM_E0_BA_E0_old}, the theoretical value of $\lambda$ is 1. Therefore, it suffices to examine the single principal value $\lambda$ to assess the consistency between the matrix obtained from $\chi_{\bm{\xi}}$ and the identity tensor $\bm{I}$. Tab. \ref{tab:lambda_chi} compares the principal value $\lambda$ for each $\chi_{\bm{\xi}}$ across different horizons $\delta$. Overall, $\chi_{\bm{\xi}}^{a}$ performs better, giving $\lambda = 1.000$ across the tested horizons, whereas $\chi_{\bm{\xi}}^{v}$ exhibits a slight positive bias ($\lambda>1$). This is consistent with the theoretical proof presented in \textbf{Appendix~A}.
	
	\begin{figure}[!htbp]
		\centering
		\includegraphics[width=0.8\textwidth]{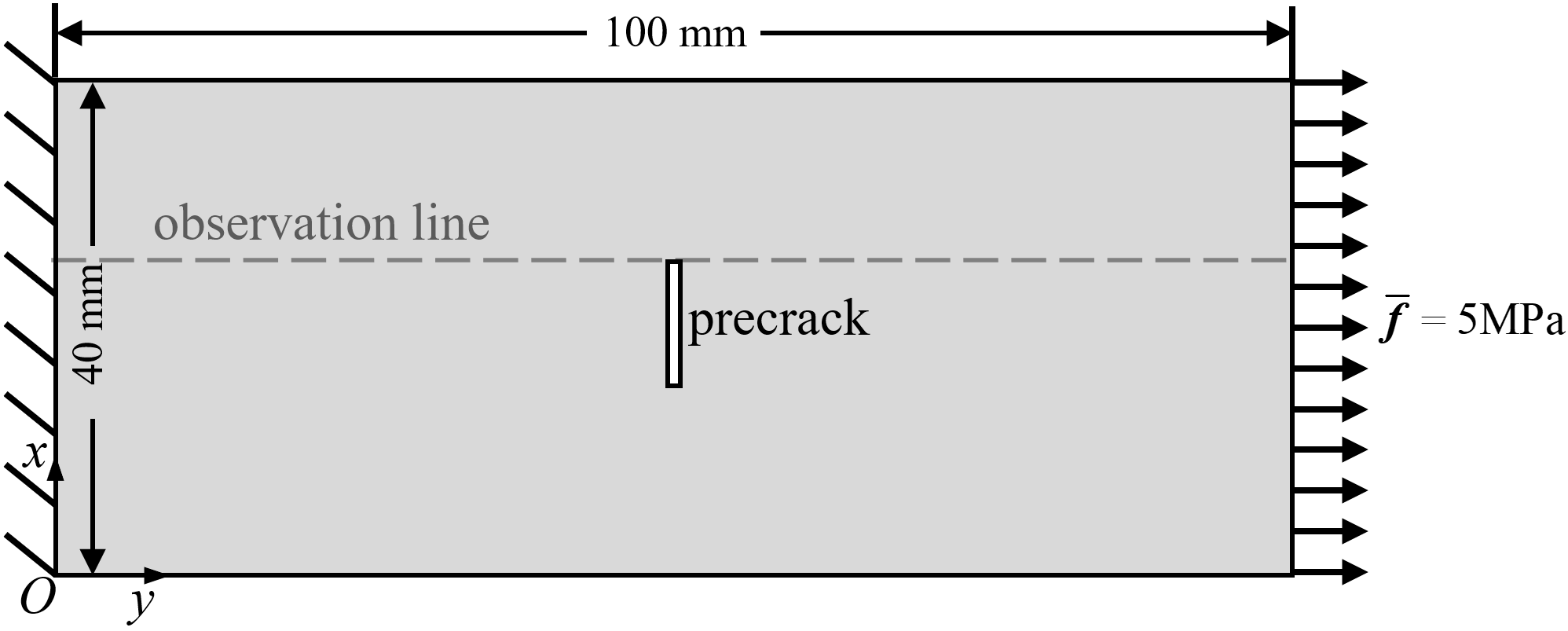}
		\caption{Geometry and boundary conditions of the rectangular plate.}
		\label{RECT_GEO}
	\end{figure}
	As shown in Fig. \ref{RECT_GEO}, a two-dimensional rectangular plate with dimensions $40\,\mathrm{mm}\times100\,\mathrm{mm}$ is considered. A horizontal precrack with a length of approximately $10\,\mathrm{mm}$ is placed at the center of the plate, extending from	$(15,50.0)\,\mathrm{mm}$ to $(25,50.0)\,\mathrm{mm}$. The lower
	boundary is fixed and a uniform traction of $5\,\mathrm{MPa}$ is applied along the upper boundary. The plate is discretized with a mesh size of	$\Delta x=0.5\,\mathrm{mm}$ and a horizon $\delta=3.03\Delta x$. An isotropic linear elastic material with $E=6\,\mathrm{GPa}$ and $\nu=0.25$ is adopted.
	\begin{figure}[!htbp]
		\centering
		\includegraphics[width=0.8\textwidth]{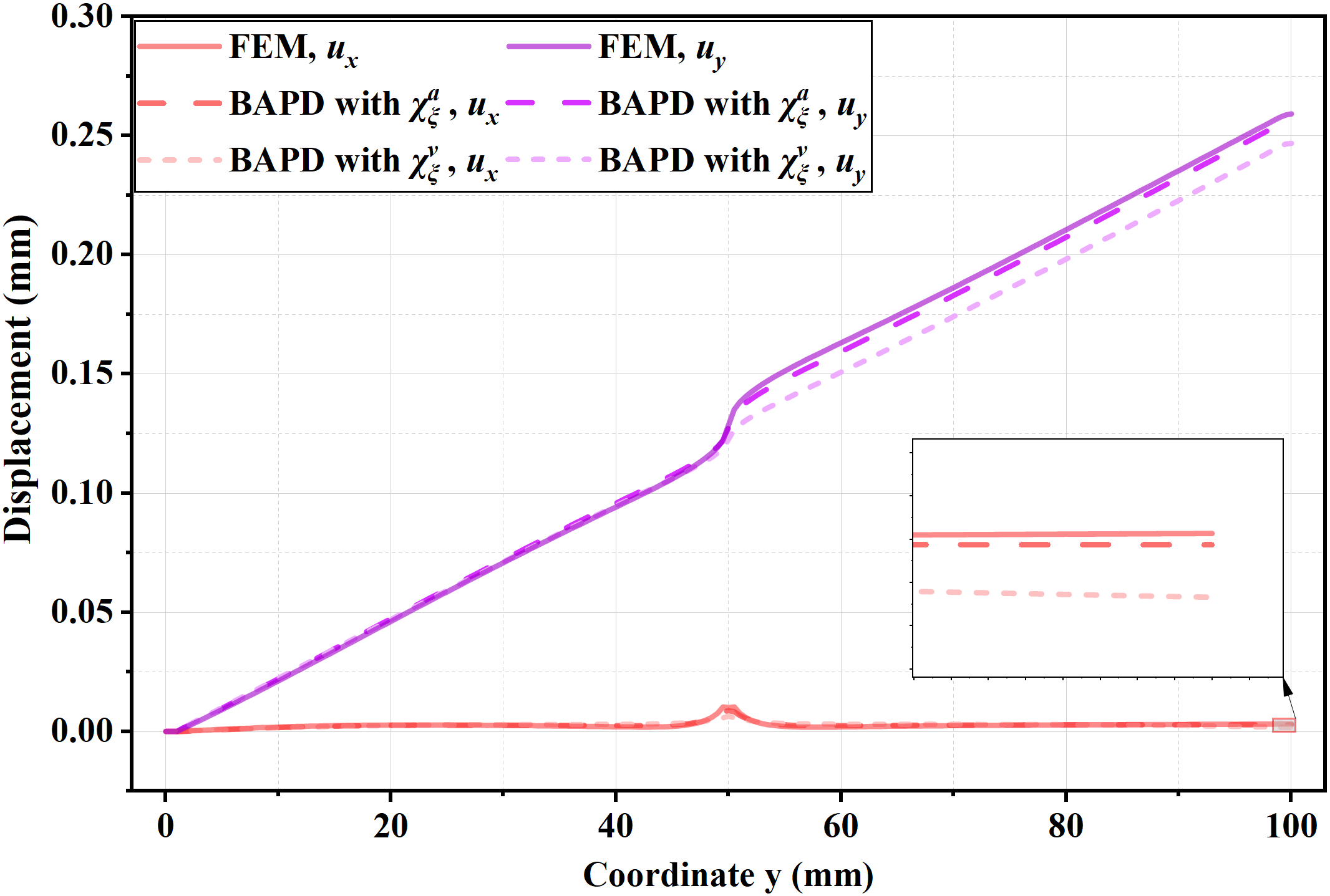}
		\caption{Comparison of the displacement $u_y$ and $u_x$ along the observation line.}
		\label{chi_num}
	\end{figure}
    
    Under quasi-static loading conditions, the rectangular plate is analyzed using the BAPD model. Two correction factors, $\chi_{\bm{\xi}}^{a}$ and $\chi_{\bm{\xi}}^{v}$, are considered. The FEM solution is adopted as the reference solution. One observation line at $x=15\,\mathrm{mm}$ is selected to compare the displacement profiles obtained from the two BAPD models with the FEM solution.
    
    Fig.~\ref{chi_num} compares the horizontal and vertical displacement profiles along the observation line. For the horizontal displacement $u_x$, the three solutions are generally close over most of the observation range. However, the enlarged view shows that the result obtained with $\chi_{\bm{\xi}}^{a}$ is closer to the FEM reference than the result obtained with $\chi_{\bm{\xi}}^{v}$. For the vertical displacement $u_y$, the two BAPD results are relatively close to the FEM solution before the precrack. After passing through the precrack, the difference between the two correction factors becomes more evident. In this domain, the result obtained with
    $\chi_{\bm{\xi}}^{a}$ remains closer to the FEM reference, whereas the result obtained with $\chi_{\bm{\xi}}^{v}$ exhibits a larger deviation.
    
    To quantify the global discrepancy, the relative $L^{2}$ norm, the relative $H^{1}$ semi-norm, and the relative energy norm are evaluated. The superscript \(\mathrm{REF}\) denotes the reference solution obtained from the FEM analysis, whereas the superscript \(\mathrm{NUM}\) denotes the numerical solution provided by the model under evaluation. The specific numerical model represented by \(\mathrm{NUM}\) is specified separately for each comparison. The relative $L^{2}$ displacement error is defined as
    
    \begin{equation}
    	\left\|e_u\right\|_{L^{2}}
    	=
    	\frac{
    		\left[
    		\int_{\Omega}
    		\left(
    		\bm{u}^{\mathrm{REF}}
    		-
    		\bm{u}^{\mathrm{NUM}}
    		\right)
    		\cdot
    		\left(
    		\bm{u}^{\mathrm{REF}}
    		-
    		\bm{u}^{\mathrm{NUM}}
    		\right)
    		\,\mathrm{d}\Omega
    		\right]^{1/2}
    	}{
    		\left[
    		\int_{\Omega}
    		\bm{u}^{\mathrm{REF}}
    		\cdot
    		\bm{u}^{\mathrm{REF}}
    		\,\mathrm{d}\Omega
    		\right]^{1/2}
    	}.
    	\label{L2}
    \end{equation}
    
    The relative $H^{1}$ semi-norm is calculated as
    
    \begin{equation}
    	\left\|e_u\right\|_{H^{1}_{\mathrm{semi}}}
    	=
    	\frac{
    		\left[
    		\int_{\Omega}
    		\left\|
    		\nabla\bm{u}^{\mathrm{REF}}
    		-
    		\nabla\bm{u}^{\mathrm{NUM}}
    		\right\|_{F}^{2}
    		\,\mathrm{d}\Omega
    		\right]^{1/2}
    	}{
    		\left[
    		\int_{\Omega}
    		\left\|
    		\nabla\bm{u}^{\mathrm{REF}}
    		\right\|_{F}^{2}
    		\,\mathrm{d}\Omega
    		\right]^{1/2}
    	}.
    	\label{H1}
    \end{equation}

    The relative energy norm is calculated as
	\begin{equation}
		\left\|e_u\right\|_{\mathrm{Energy}}
		=
		\left[
		\frac{
			\displaystyle
			\int_{\Omega}
			\left[
			W^{\mathrm{REF}}(\bm{x})
			-
			W^{\mathrm{NUM}}(\bm{x})
			\right]^2
			\,\mathrm{d}\Omega
		}{
			\displaystyle
			\int_{\Omega}
			\left[
			W^{\mathrm{REF}}(\bm{x})
			\right]^2
			\,\mathrm{d}\Omega
		}
		\right]^{1/2},
		\label{energy}
	\end{equation}
    where \(W^{\mathrm{REF}}(\bm{x})\) and \(W^{\mathrm{NUM}}(\bm{x})\) denote the strain energy densities obtained from the reference and numerical solutions, respectively. 

    The superscript \(\mathrm{NUM}\) refers to the BAPD solution obtained with different correction factors. The calculated relative norms are summarized in Tab.~\ref{tab:error_norms}. The results show that the errors obtained with $\chi_{\bm{\xi}}^{a}$ are smaller than those obtained with $\chi_{\bm{\xi}}^{v}$ for all three error measures.
    
    \begin{table}[htbp]
    	\centering
    	\caption{Relative error norms obtained with different correction
    		factors.}
    	\label{tab:error_norms}
    	\begin{tabular}{c c c c}
    		\toprule
    		Correction factor
    		& $L^{2}$ norm (\%)
    		& $H^{1}$ semi-norm (\%)
    		& Energy norm (\%)\\
    		\midrule
    		$\chi_{\bm{\xi}}^{a}$
    		& 1.5574
    		& 1.8124
    		& 1.8135 \\
    		$\chi_{\bm{\xi}}^{v}$
    		& 5.7974
    		& 6.2821
    		& 6.2494 \\
    		\bottomrule
    	\end{tabular}
    \end{table}
    
    In percentage form, the corresponding errors are $1.5574\%$, $1.8124\%$, and $1.8135\%$ for $\chi_{\bm{\xi}}^{a}$, and $5.7974\%$, $6.2821\%$, and $6.2494\%$ for
    $\chi_{\bm{\xi}}^{v}$, respectively. Therefore, under the present case, $\chi_{\bm{\xi}}^{a}$ produces smaller global displacement and energy errors than $\chi_{\bm{\xi}}^{v}$.

	\paragraph{\textbf{Remark}}
		 In Tab.~\ref{tab:lambda_chi}, \(\chi^{v}_{\bm\xi}\) often gives \(\lambda>1\), so the tensor term in Eq.~(\ref{E_CCM}) does not reduce to \(\bm I\) as \(\alpha(\bm x,t),\alpha(\bm x',t)\to 1\) (see Eq.~(\ref{CCM_E0_BA_E0_old})). This may lead to a negative CCM stiffness tensor inside the BAPD domain.

	\subsection{Evaluation of the BAPD-CCM model under quasi-static loading}

	\begin{figure}[H]
		\centering
		\includegraphics[width=\textwidth]{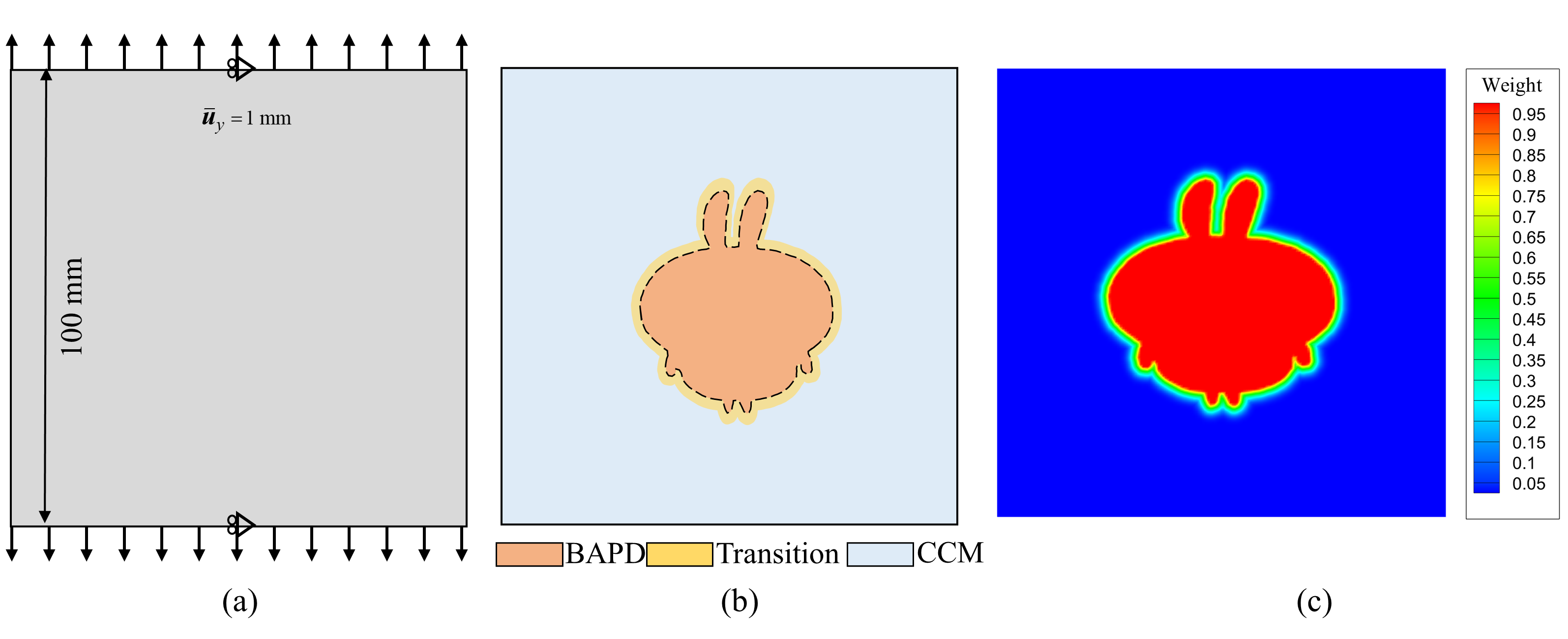}
		\caption{Geometry and boundary conditions (a), partition into CCM, transition, and
			BAPD domains (b), and distribution of the ``Morphing" function $\alpha$ (c).}
		\label{plane_geo}
	\end{figure}
	A $100\,\mathrm{mm}$ square plate subjected to quasi-static uniaxial tension is considered to evaluate the accuracy of the proposed BAPD-CCM model. The geometry and boundary conditions are shown in Fig. \ref{plane_geo} (a). The plate has a side length of $100\,\mathrm{mm}$. Vertical displacements $u_y = \pm 1 \ \mathrm{mm}$ are applied on the top and bottom edges, and horizontal displacements $u_x = 0 \ \mathrm{mm}$ are fixed at the midpoints of these edges. The mesh size is set to $\Delta x=0.5\,\mathrm{mm}$, yielding $40{,}000$ quadrilateral elements and $40{,}401$ nodes. The horizon is chosen as $\delta=3.03\,\Delta x$, and a constant influence function is adopted. The elastic modulus and Poisson's ratio of the material are $E = 72 \ \mathrm{GPa}$ and $\nu = 1/3$, respectively. As shown in Fig. \ref{plane_geo} (b), a rabbit-shaped BAPD domain is embedded at the plate center, with a transition domain of width $6 \Delta x$. Fig. \ref{plane_geo} (c) illustrates the distribution of the ``Morphing" function $\alpha$.
   	\begin{figure}[H]
		\centering
		\includegraphics[width=\textwidth]{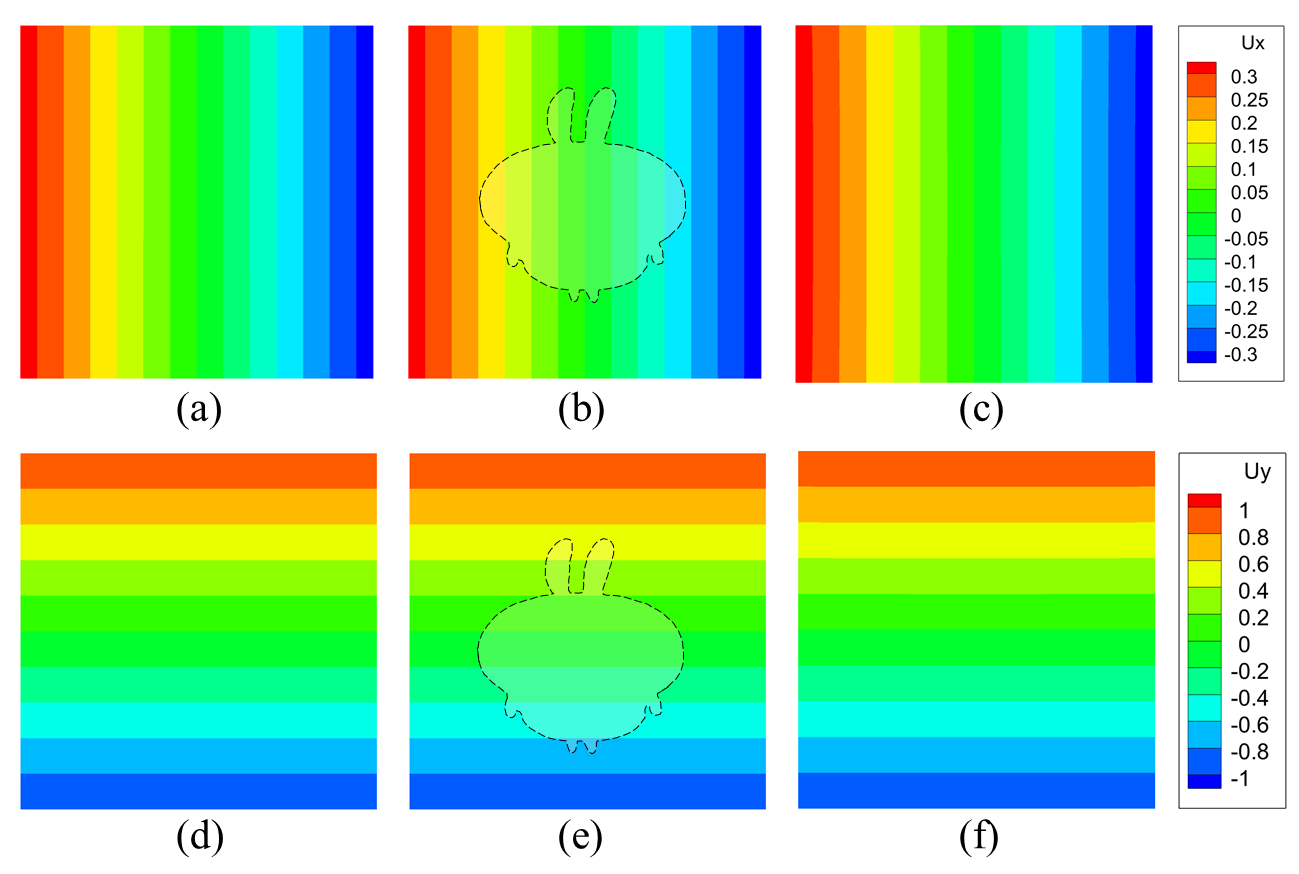}
		\caption{Displacement contours under quasi-static loading: $u_x$ from FEM (a),	BAPD-CCM (b), and BAPD (c); $u_y$ from FEM (d), BAPD-CCM (e), and BAPD (f).}
		\label{plane_dis}
	\end{figure}
	
	Fig. \ref{plane_dis} shows displacement contours obtained from the FEM, the BAPD-CCM model and the BAPD model. All three solutions exhibit broadly uniform displacement fields; however, the BAPD solutions still show noticeably curved contours near the plate corners, especially in Fig.~\ref{plane_dis} (c), indicating boundary effects, whereas the BAPD-CCM model does not. 
	
	\begin{figure}[!htbp]
		\centering
		\includegraphics[width=0.9\textwidth]{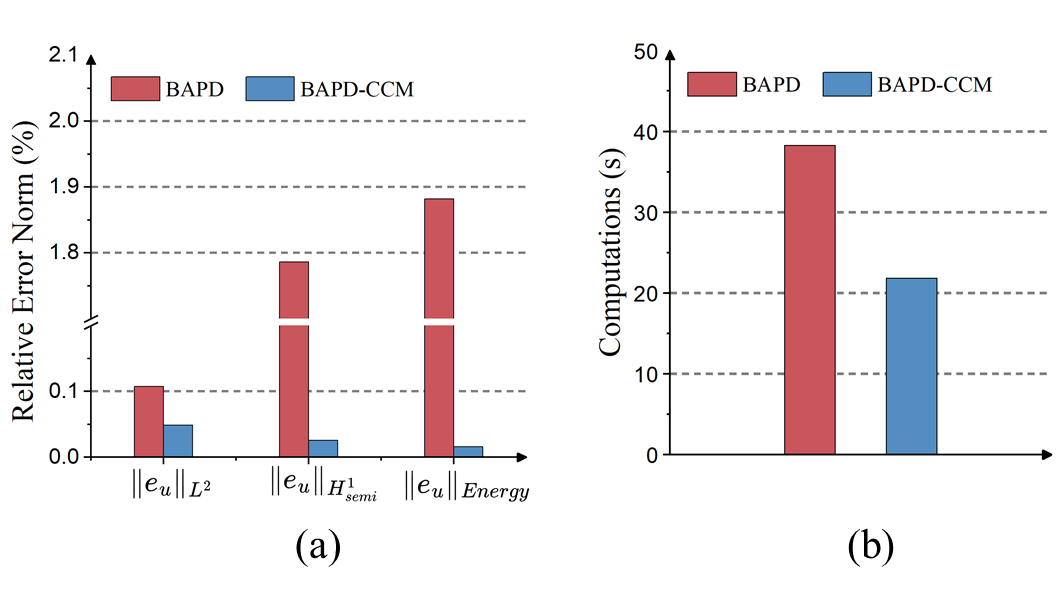}
		\caption{Comparison between the BAPD model and the BAPD-CCM model under quasi-static loading: relative error norms (a) and computation time per load step (b).}
		\label{static_errornorm_caltime}
	\end{figure}

	To quantify the global error, we evaluate the norms in Eq. \eqref{L2}, \eqref{H1} and \eqref{energy} using the FEM solution as the reference, and the superscript \(\mathrm{NUM}\) refers to the BAPD-CCM solution. As shown in Fig. \ref{static_errornorm_caltime} (a), the relative errors $\Arrowvert e_{u} \Arrowvert_{L^{2}} , \Arrowvert e_{u} \Arrowvert_{H^{1}_{\mathrm{semi}}}$, and $\Arrowvert e_{u} \Arrowvert_{Energy}$ of BAPD are $0.108\%$, $1.786\%$, and $1.882\%$, respectively, whereas for BAPD-CCM the corresponding errors are $0.049\%$, $0.026\%$, and $0.016\%$, which are markedly smaller. In terms of efficiency for a single load step (Fig. \ref{static_errornorm_caltime} (b)), BAPD takes $38.307\,\mathrm{s}$ and BAPD-CCM takes $21.857\,\mathrm{s}$, corresponding to a speed-up of approximately $43\%$.
	
	\subsection{Evaluation of the BAPD-CCM model under dynamic loading}
	This subsection validates the BAPD-CCM model for a dynamic loading case, building on the setup introduced in the previous subsection. The geometry and BAPD domain are identical to the quasi-static case; the elastic modulus, Poisson's ratio and density of the material are $E = 72~\mathrm{GPa}$, $\nu = 1/3$ and $\rho = 2{,}450 ~ \mathrm{kg/m^{3}}$, respectively. The bottom edge is fixed in the \(y\)-direction, the midpoint of the bottom edge is constrained in \(x\) to suppress rigid body motion, and a traction boundary condition \(\bar{f}=1~\mathrm{GPa}\) is applied to the top edge. The mesh size is $\Delta x = 1 ~ \mathrm{mm}$, yielding $10{,}000$ elements and $10{,}201$ nodes. For the dynamic analysis, the time step is set to \(\Delta t=5~\times 10^{-9}\mathrm{s}\), and \(50{,}000\) steps are performed, corresponding to a total duration of \(0.25~ \mathrm{ms}\).
	\begin{figure}[H]
		\centering
		\includegraphics[width=0.8\textwidth]{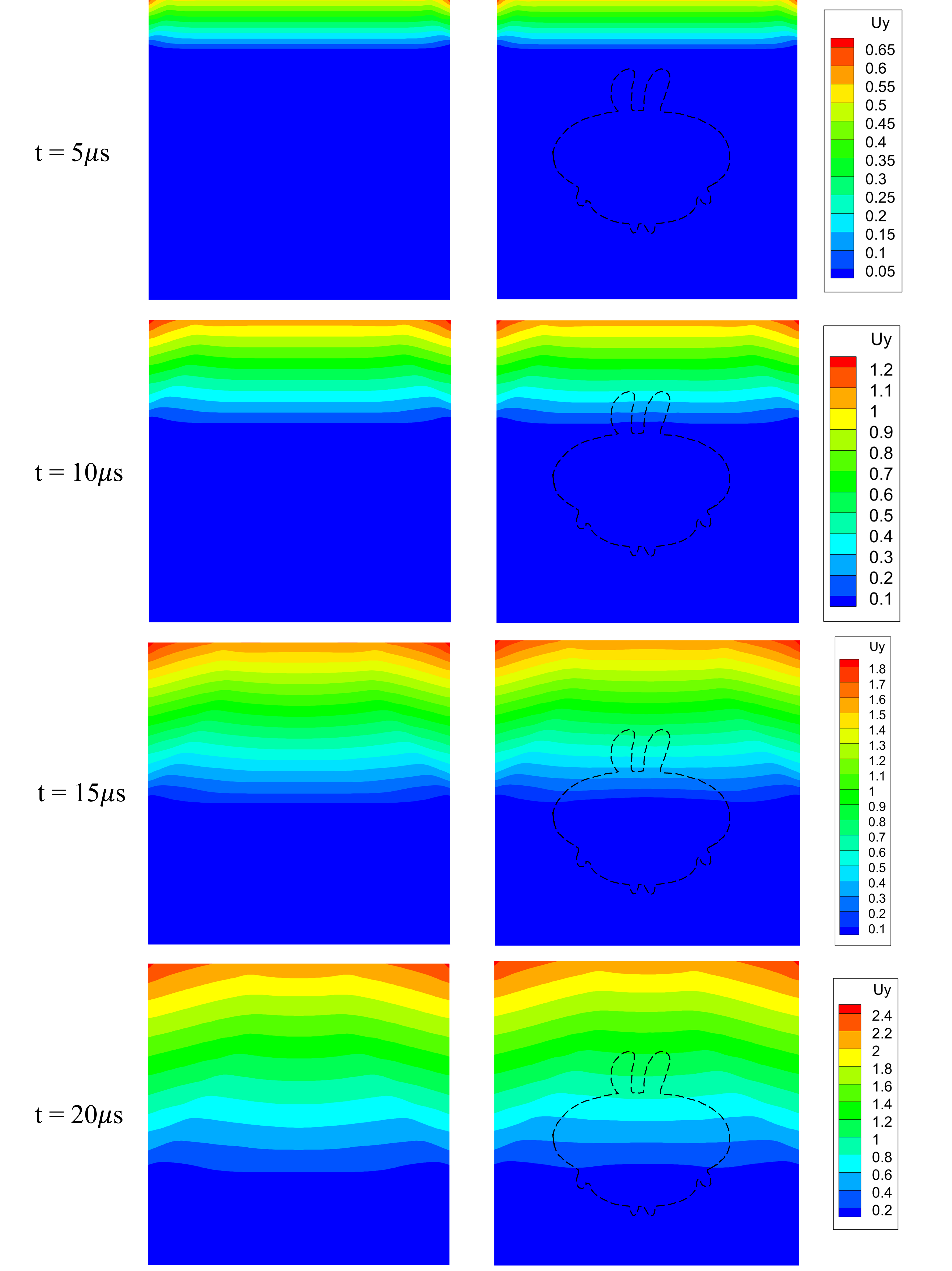}
		\caption{Comparison of the y-direction displacement $u_{y}$ contours of FEM (left) and BAPD-CCM (right) at different times.}
		\label{test_3_dy_uy}
	\end{figure}
	\begin{figure}[H]
		\centering
		\includegraphics[width=\textwidth]{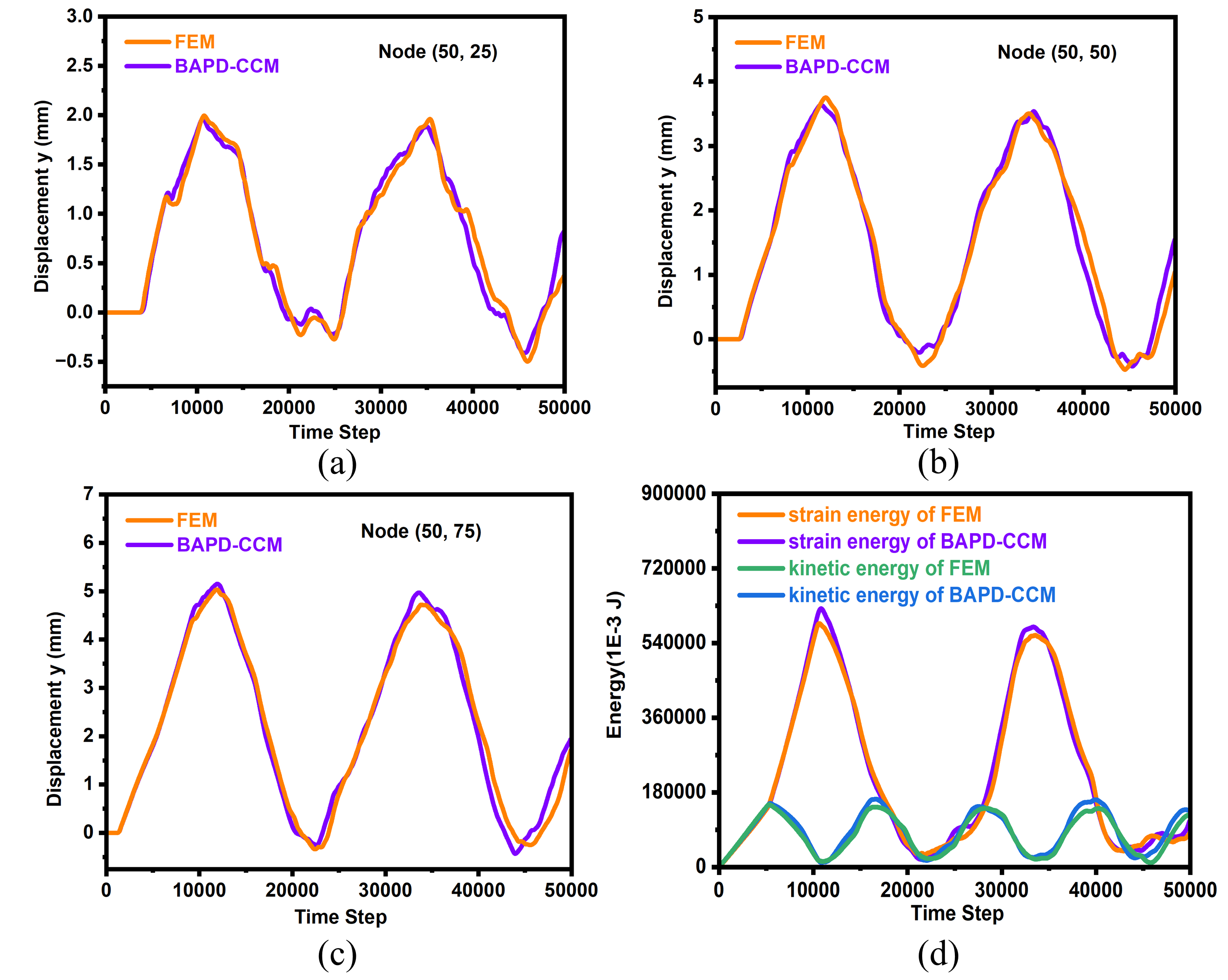}
		\caption{Dynamic response comparison between BAPD-CCM and FEM: vertical
			displacement $u_{y}$ at $\text{Node} (x=50~\mathrm{mm},y=25~\mathrm{mm})$ (a),
			$\text{Node} (x=50~\mathrm{mm},y=50~\mathrm{mm})$ (b), and
			$\text{Node} (x=50~\mathrm{mm},y=75~\mathrm{mm})$ (c); strain and kinetic energy
			histories (d).}
		\label{DY_dis_ene}
	\end{figure}
    To clearly observe the propagation of the displacement wave, Fig. \ref{test_3_dy_uy} presents a comparison of the $u_y$ contours obtained by FEM and the BAPD-CCM model within the first $20\,\mathrm{\mu s}$. It can be seen that the two displacement fields are in close agreement in both spatial distribution and magnitude. However, the BAPD-CCM model exhibits a slight bias at the coupling interface. In addition, three representative nodes at \(\text{Node} \  (x=50~\mathrm{mm},y=25~\mathrm{mm})\), \(\text{Node} \ (x=50~\mathrm{mm},y=50~\mathrm{mm})\), and \(\text{Node} \ (x=50~\mathrm{mm},y=75~\mathrm{mm})\) are monitored. Fig.~\ref{DY_dis_ene} (a-c) compares the full time results of the vertical displacement \(u_y\) from BAPD-CCM and FEM. The maximum deviation between the BAPD-CCM and FEM displacement curves is below $3\%$. The strain energy and kinetic energy are also compared over all time steps; as shown in Fig.~\ref{DY_dis_ene} (d), both the trends and magnitudes are similar.

	\section{Numerical examples}\label{S7}
	This section investigates three two- and three-dimensional dynamic and quasi-static cases to demonstrate the capability of the proposed model in handling fracture problems.
	\subsection{Mode \uppercase\expandafter{\romannumeral1} fracture under quasi-static loading}
   	\begin{figure}[!htbp]
		\centering
		\includegraphics[width=0.8\textwidth]{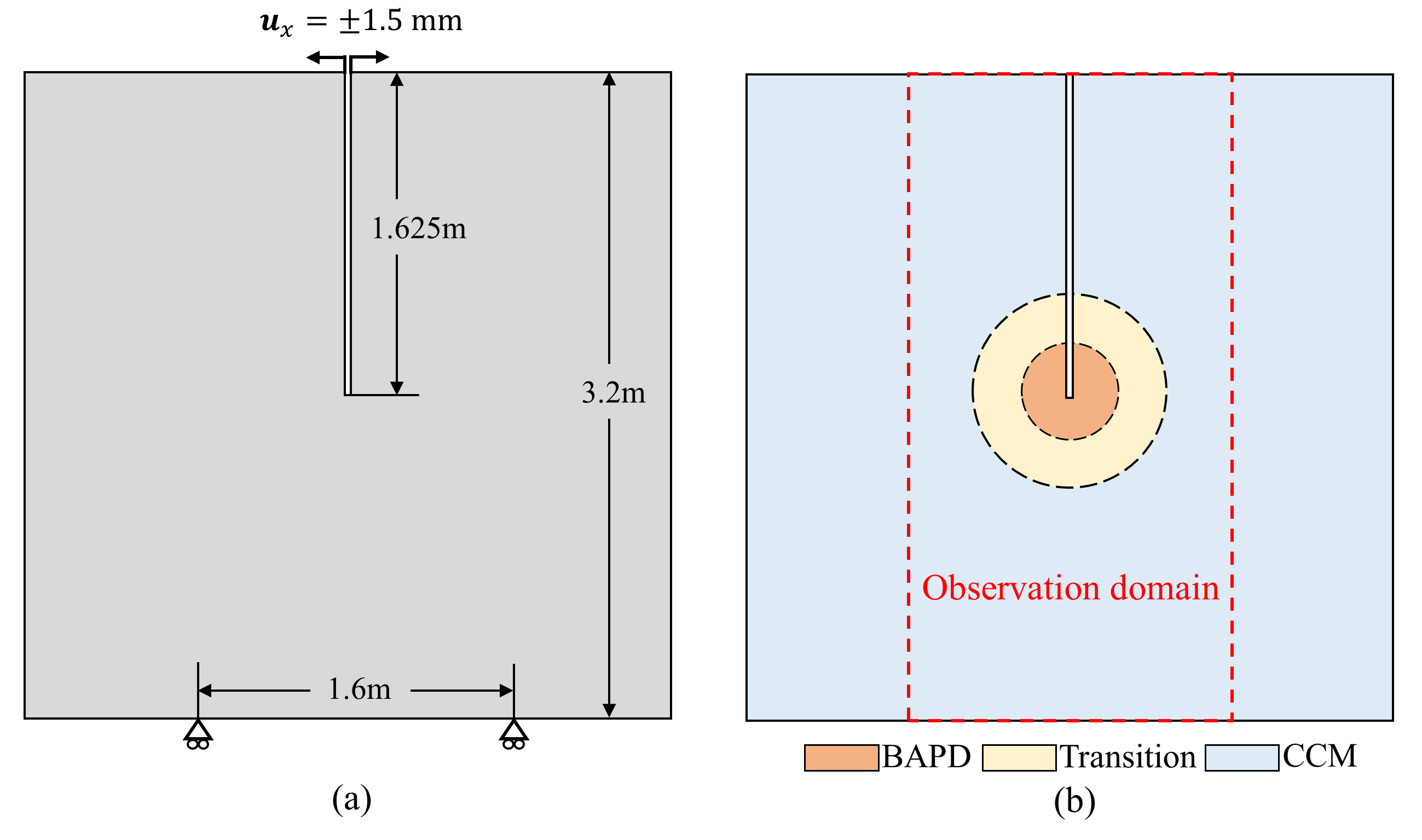}
		\caption{The geometry, boundary conditions (a), and the partitioning in the BAPD-CCM model (b) of the Mode \uppercase\expandafter{\romannumeral1} fracture.}
		\label{MODE_1_GEO}
	\end{figure}
	A square plate with a side length of $3.2~\mathrm{m}$ containing a $1.625~\mathrm{m}$ long precrack centered along the upper edge, as illustrated in Fig.~\ref{MODE_1_GEO} (a) \cite{Trunk1999}. The plate material is isotropic and linear elastic with an elastic modulus of $28.3~\mathrm{GPa}$ and a Poisson's ratio of $0.3$. The material fracture energy $G_c$ is $196~\mathrm{J/m^2}$. The horizon $\delta$ is set as $3.03\Delta x$. The domain is discretized into a uniform mesh with 26,002 nodes and 25,600 elements, corresponding to a mesh size of $\Delta x = 20~\mathrm{mm}$. The problem is analyzed using the BAPD-CCM model with bond-breaking-induced criterion, in which the BAPD domain is assigned around the crack tip, as depicted in Fig.~\ref{MODE_1_GEO} (b). The displacement loading is applied in 100 incremental steps.
   	\begin{figure}[!htbp]
		\centering
		\includegraphics[width=0.85\textwidth]{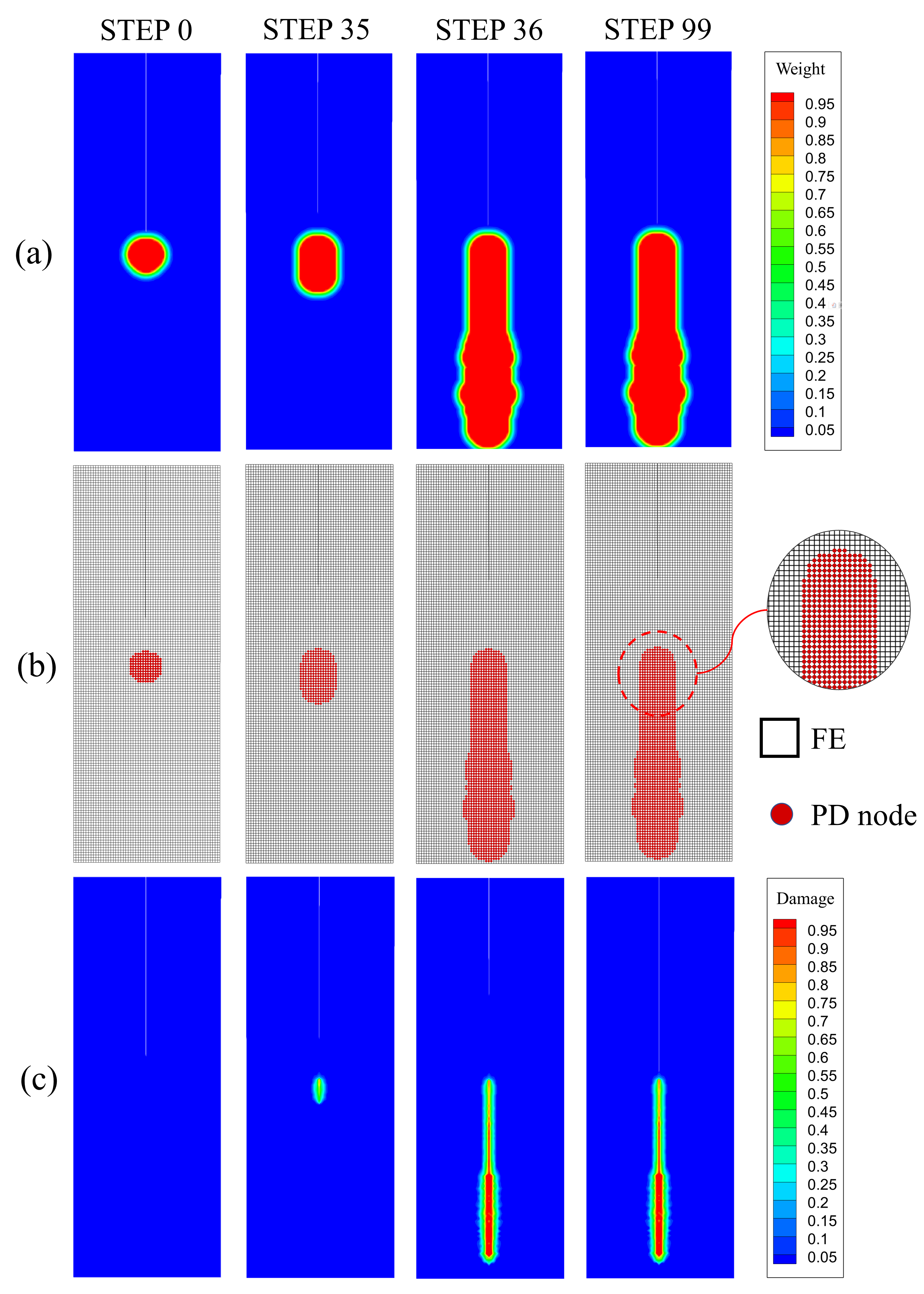}
		\caption{The evolution of $\alpha$ value (a), elements and particles (b), and damage contour (c) of different incremental steps.}
		\label{case_1_wei_dam}
	\end{figure}
    
    Fig.~\ref{case_1_wei_dam} illustrates the evolution of the $\alpha$ value, the element/PD particle topology, and the propagation of the damaged domain at selected load steps within the observation domain. Since damage develops only below the crack tip, only a $1.2~\mathrm{m}$-wide portion in the middle of the plate is shown. At the 35th load step, little damage has occurred yet, whereas at the 36th step several bonds near the crack tip start to break. As bond breakage accumulates, the region where $\alpha > 0$ gradually expands outward (Fig.~\ref{case_1_wei_dam} (a)), the nodes of element in this domain are rapidly converted into PD particles (Fig.~\ref{case_1_wei_dam} (b)), and the damaged domain grows quickly downward along the crack path (Fig.~\ref{case_1_wei_dam} (c)). Afterwards, up to the 99th step, the damage front advances only slightly, indicating that once crack initiation takes place, the main crack extension is completed within a short time, which is consistent with the characteristics of brittle fracture.
    
    \begin{figure}[!htbp]
    	\centering
    	\includegraphics[width=0.7\textwidth]{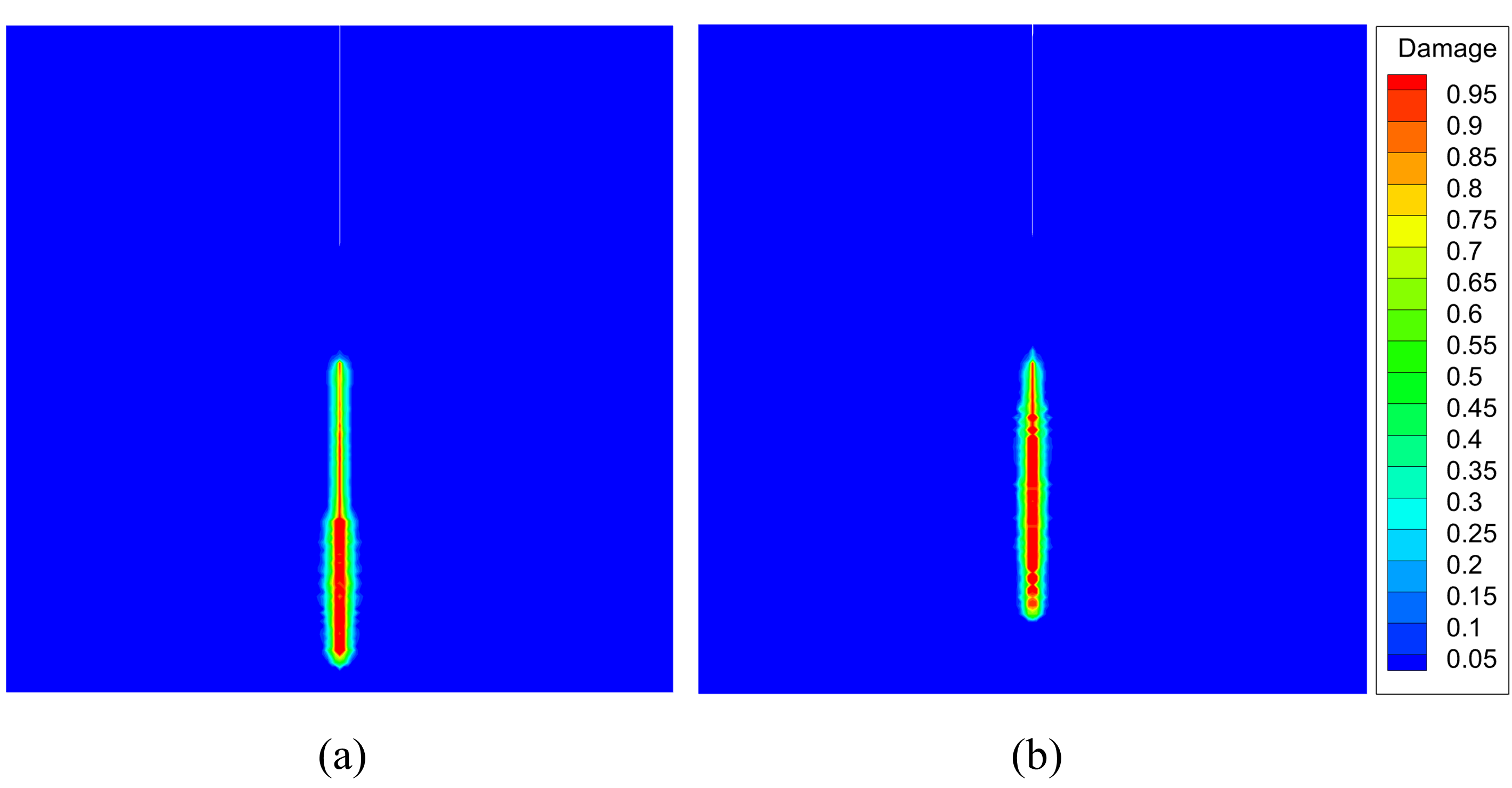}
    	\caption{The damage contours of BAPD-CCM model (a) and BAPD model (b).}
    	\label{CASE_1_DAM}
    \end{figure}

    \begin{figure}[!htbp]
		\centering
		\includegraphics[width=\textwidth]{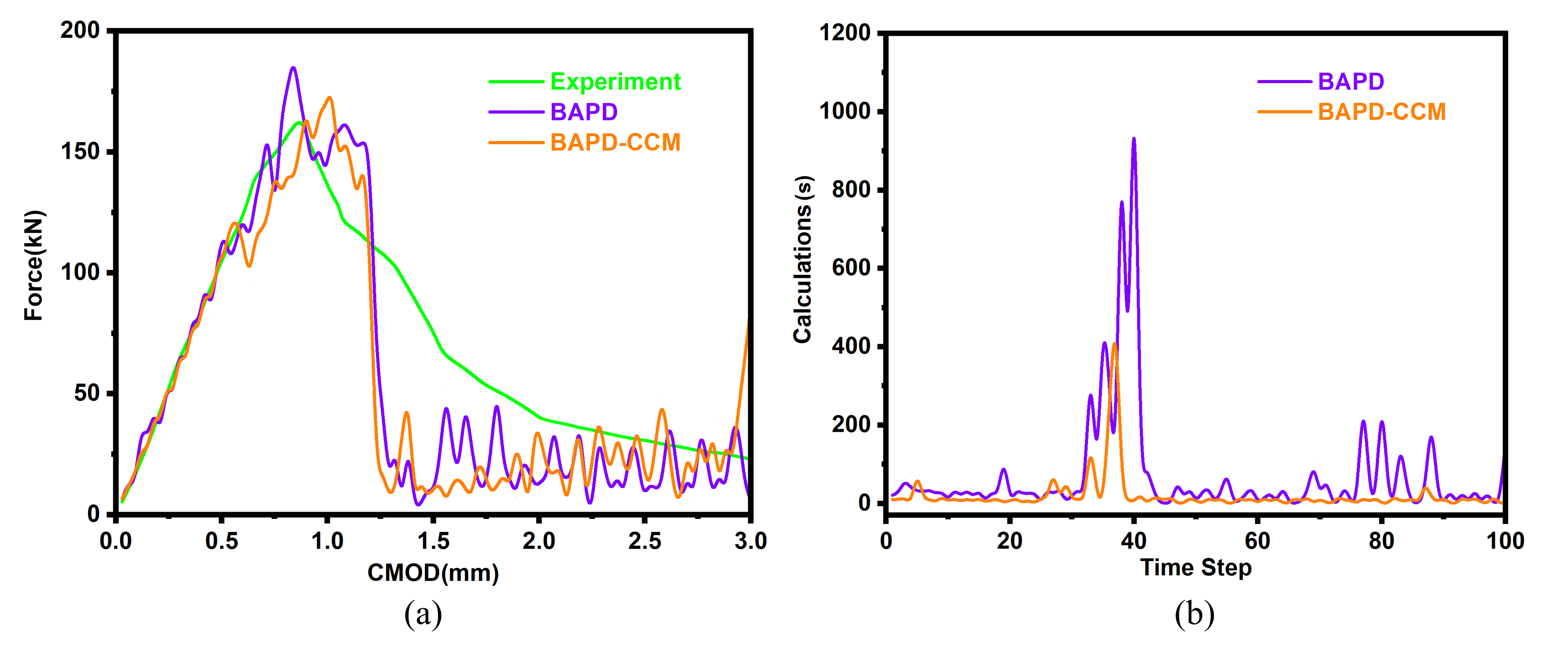}
		\caption{Comparison of the force-CMOD curves and calculation time.}
		\label{CASE_1_FORCE_CALCULATIONS}
	\end{figure}

    Fig. \ref{CASE_1_DAM} compares the final damage contours of the BAPD-CCM and BAPD models. While the crack paths are generally consistent, the damage-band width differs. A possible reason is that, in the bond-breaking-induced algorithm, the BAPD domain is activated only in the vicinity of newly broken bonds and then expands progressively with crack growth. This confines damage evolution to a narrow region around the crack path, leading to a more slender crack pattern. The force-CMOD curves show that both BAPD and BAPD-CCM models capture the overall experimental trend (Fig.~\ref{CASE_1_FORCE_CALCULATIONS} (a)); the peak load coincides with the initiation of large-scale fracture, after which a significant decrease in load is observed. The computational-time comparison indicates that BAPD-CCM significantly reduces the cost (Fig.~\ref{CASE_1_FORCE_CALCULATIONS} (b)), reducing the total simulation time by 72.144\%.
    
    \subsection{Dynamic crack branching in brittle materials}
    \begin{figure}[H]
    	\centering
    	\includegraphics[width=0.7\textwidth]{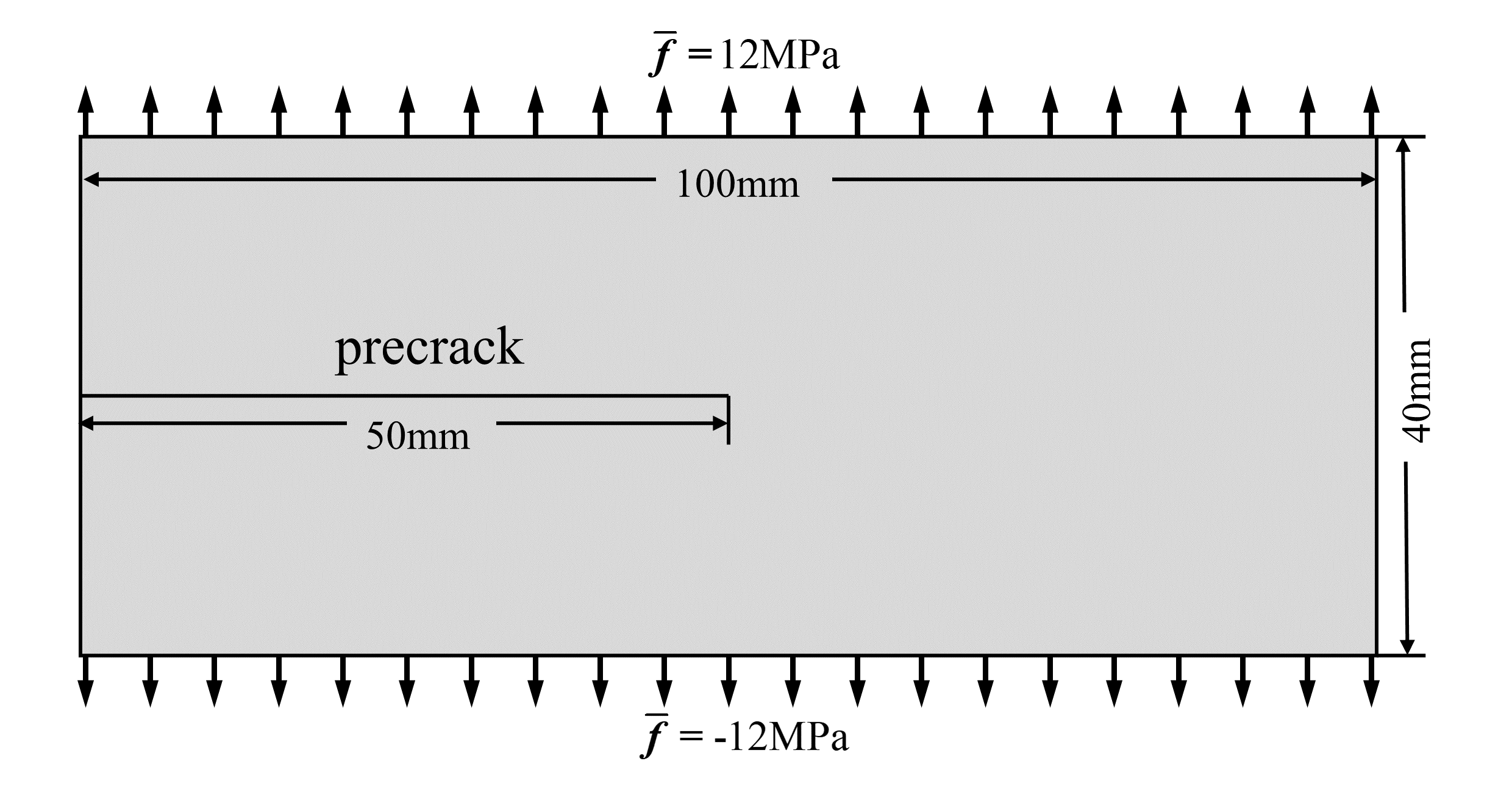}
    	\caption{The geometry and boundary conditions of the crack branching case.}
    	\label{CASE_2_GEO}
    \end{figure}
    \begin{figure}[H]
    	\centering
    	\includegraphics[width=0.9\textwidth]{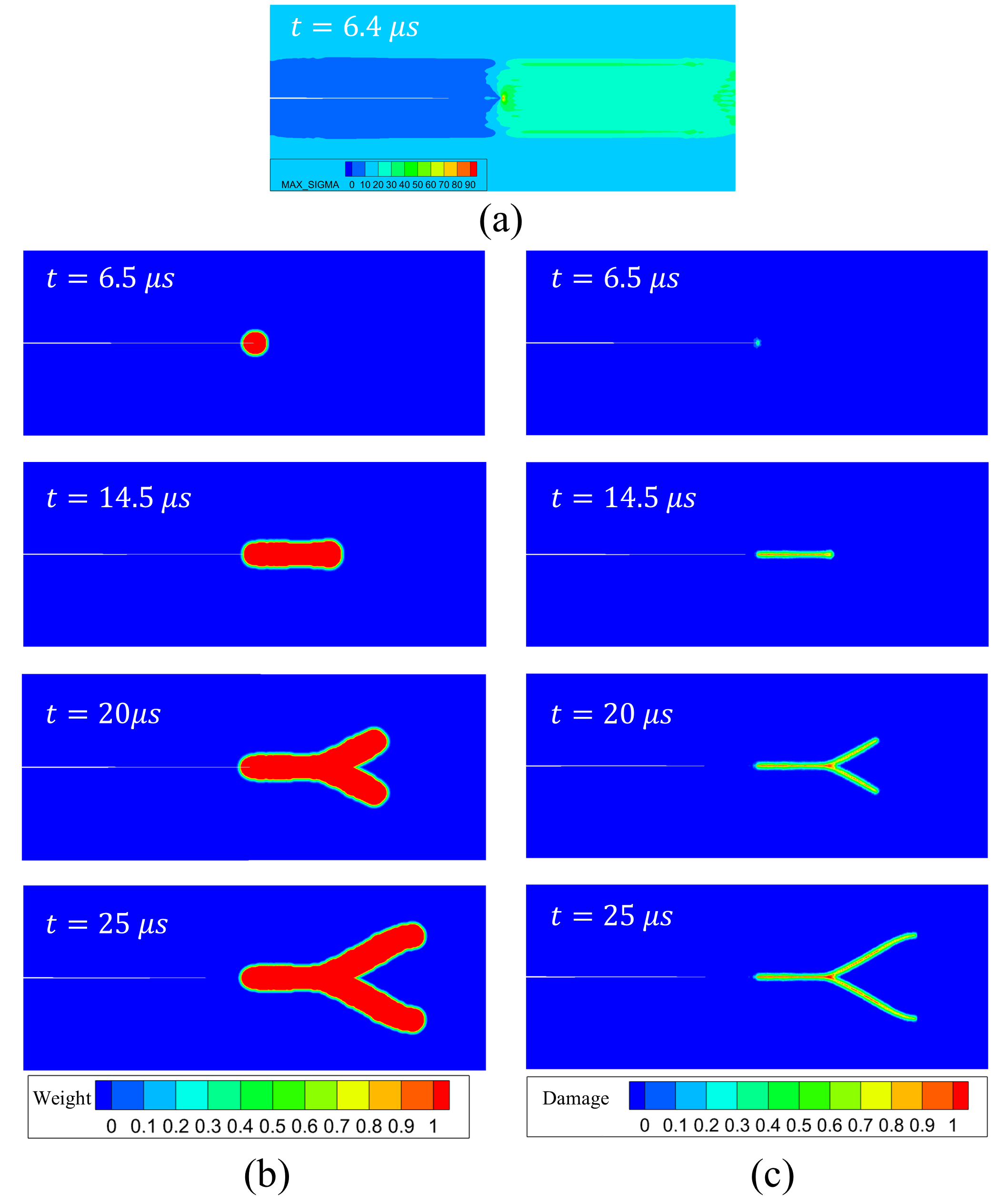}
    	\caption{The maximum principal stress contour at $6.5 \mathrm{\mu s}$ (a), the distribution of $\alpha$ value (b) and damage contour (c) at selected time instants. }
    	\label{CASE_2_DAM_WEI}
    \end{figure}
    This subsection revisits a classical dynamic crack branching benchmark. The specimen is a slender rectangular plate of length $100~\mathrm{mm}$ and width $40~\mathrm{mm}$ containing a $50~\mathrm{mm}$ precrack along the midline, as shown in Fig.~\ref{CASE_2_GEO}. The domain is discretized into $64{,}000$ uniform quadrilateral elements with a mesh size of $\Delta x = 0.25~\mathrm{mm}$, and the horizon is chosen as $\delta = 4.04\,\Delta x$. The material (soda-lime glass) is assumed to be brittle, homogeneous and isotropic linear elastic with density $\rho = 2.44\times10^{3}~\mathrm{kg/m^{3}}$, elastic modulus $E = 72~\mathrm{GPa}$ and Poisson's ratio $\nu = 1/3$, and the material fracture energy is $G_c = 135~\mathrm{J/m^2}$. Uniform normal tractions $\bar{f}=12~\mathrm{MPa}$ are applied to the top and bottom edges of the plate and then kept constant for a total loading duration of $30~\mathrm{\mu s}$, which is solved using $30{,}000$ time steps. The problem is analyzed using the BAPD-CCM model, and the strength-induced criterion is adopted to activate the BAPD domain.

 	\begin{figure}[h]
		\centering
		\includegraphics[width=0.7\textwidth]{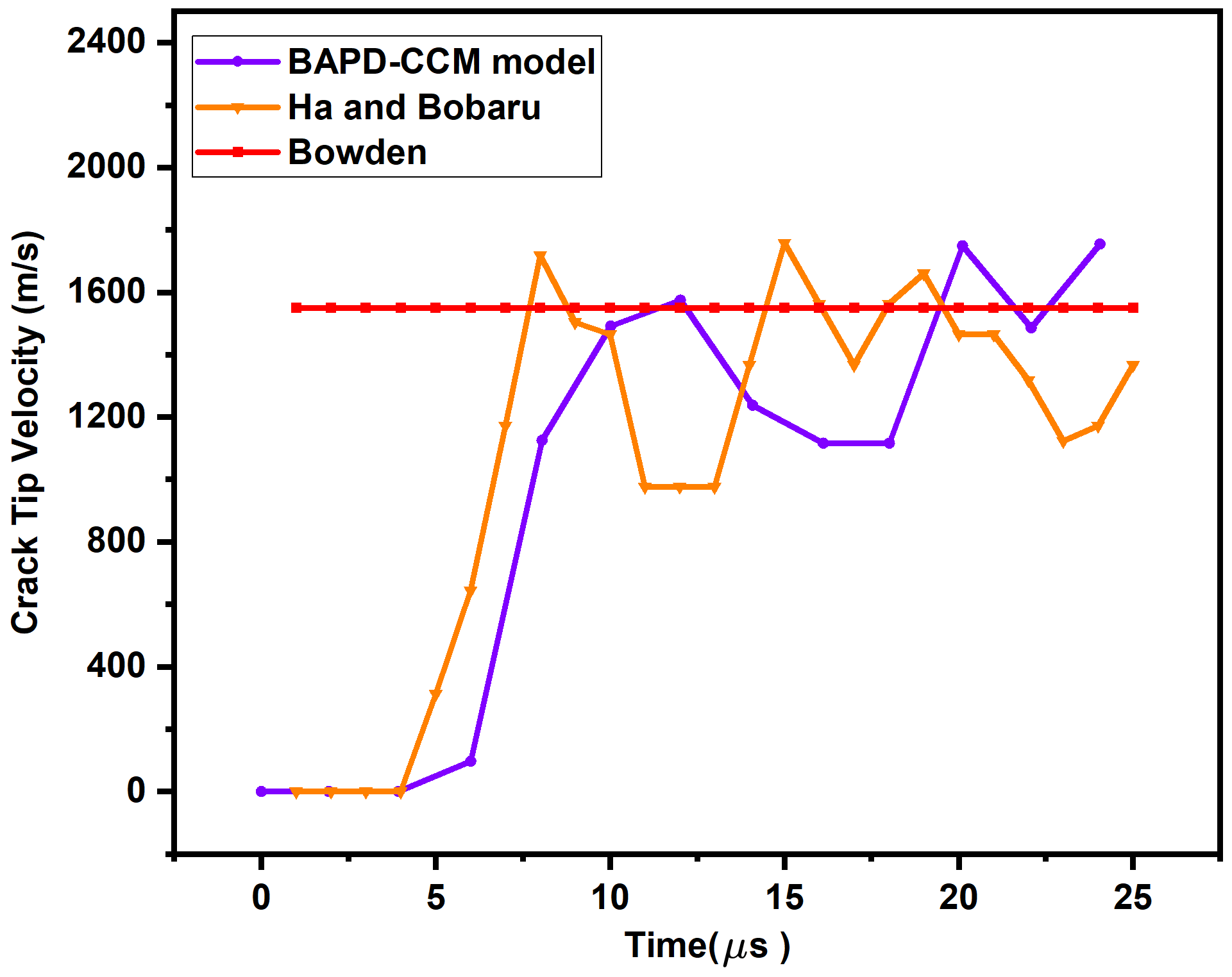}
		\caption{Crack-tip velocity predicted by the present BAPD-CCM model, compared with the experimental data of Bowden et al. \cite{1968bosu} and the numerical results of Ha and Bobaru \cite{Ha2011_BrittlePD}.}
		\label{CRACK_VEL}
	\end{figure}

	Fig. \ref{CASE_2_DAM_WEI} presents the evolution of dynamic fracture in this problem. At \(t = 6.4~\mathrm{\mu s}\), the stress waves generated from the top and bottom boundaries meet in the middle of the plate and produce a pronounced stress concentration at the precrack tip (Fig.~\ref{CASE_2_DAM_WEI} (a)). At this instant, some material points first satisfy the maximum principal stress criterion and fail, and the BAPD domain is adaptively induced in the vicinity of the crack tip, where damage starts to nucleate. Subsequently, the BAPD domain propagates to the right following the advancing crack tip, while the damage continues to grow within this domain. From \(t = 14.5~\mathrm{\mu s}\) to \(t = 20~\mathrm{\mu s}\), the damage begins to exhibit a branching pattern, and the BAPD domain expands upward and downward accordingly. A possible reason is that the interaction between the reflected stress wave and the subsequent incident wave near the crack tip gives rise to two symmetric off-axis stress concentration domains, which in turn promote crack branching. In addition, the crack tip velocity was evaluated, and the predicted values are in close agreement with experimental results \cite{1968bosu} and previous numerical simulations \cite{Ha2011_BrittlePD} (Fig. \ref{CRACK_VEL}).

	\subsection{Concrete slab subjected to drop-weight impact}

\begin{figure}[!htbp]
		\centering
		\includegraphics[width=0.7\textwidth]{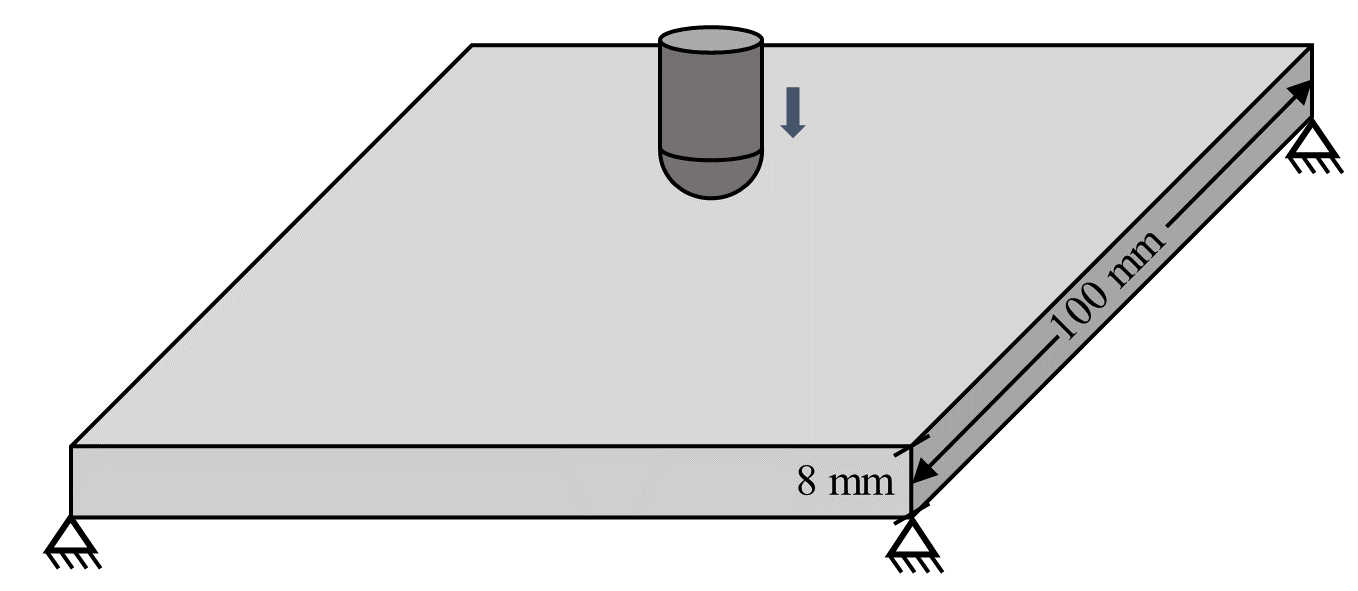}
		\caption{Schematic illustration of concrete subjected to drop-weight impact.}
		\label{3D_GEO}
	\end{figure}

\begin{figure}[!htbp]
		\centering
		\includegraphics[width=\textwidth]{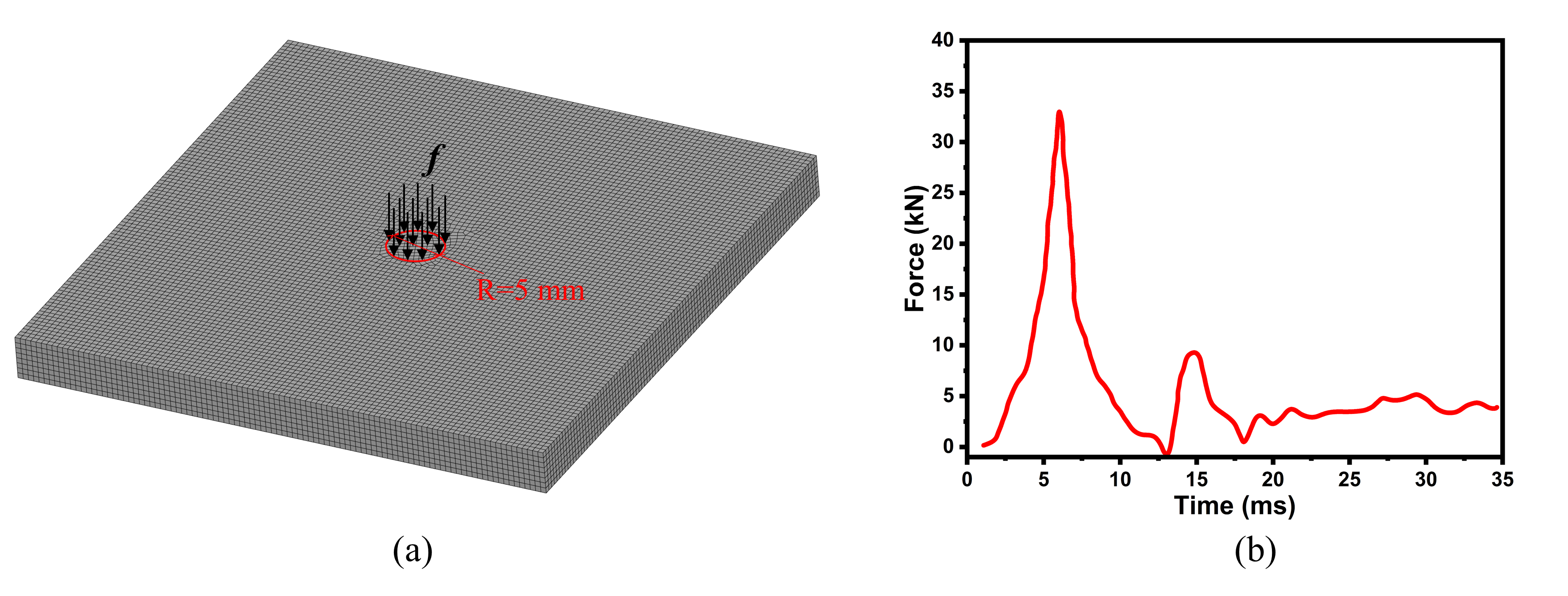}
		\caption{Simplified schematic of the drop-weight impact load (a) and the corresponding experimental force-time history \cite{LOW_IMPACT} (b).}
		\label{3D_MESH_FORCE}
	\end{figure}

	A square concrete plate subjected to drop-weight impact \cite{LOW_IMPACT} is investigated, as shown in Fig.~\ref{3D_GEO}. The experimental setup is simplified to the configuration in Fig.~\ref{3D_MESH_FORCE}(a), where the applied load is prescribed directly from the experimental data \cite{LOW_IMPACT}. The load application radius is ramped from $0$ to $5~\mathrm{mm}$ within the first $5~\mathrm{ms}$. The mesh size $\Delta x$ is $1~\mathrm{mm}$, resulting in 80{,}104 hexahedral elements and 91{,}926 nodes. The horizon is set to 3.03$\Delta x$. The material density and Young's modulus are taken as $\rho = 2.5\times10^{3}~\mathrm{kg/m^{3}}$ and $E = 30~\mathrm{GPa}$, respectively, with Poisson's ratio $\nu=0.2$, and the critical stretch is set to $1.0\times10^{-4}$. The total calculation time is $37.4~\mathrm{ms}$, and a total of 500{,}000 time steps are performed. The problem is analyzed using the BAPD-CCM model, and the strength-induced criterion is used to induce the BAPD domain.
	
 	\begin{figure}[!htbp]
		\centering
		\includegraphics[width=\textwidth]{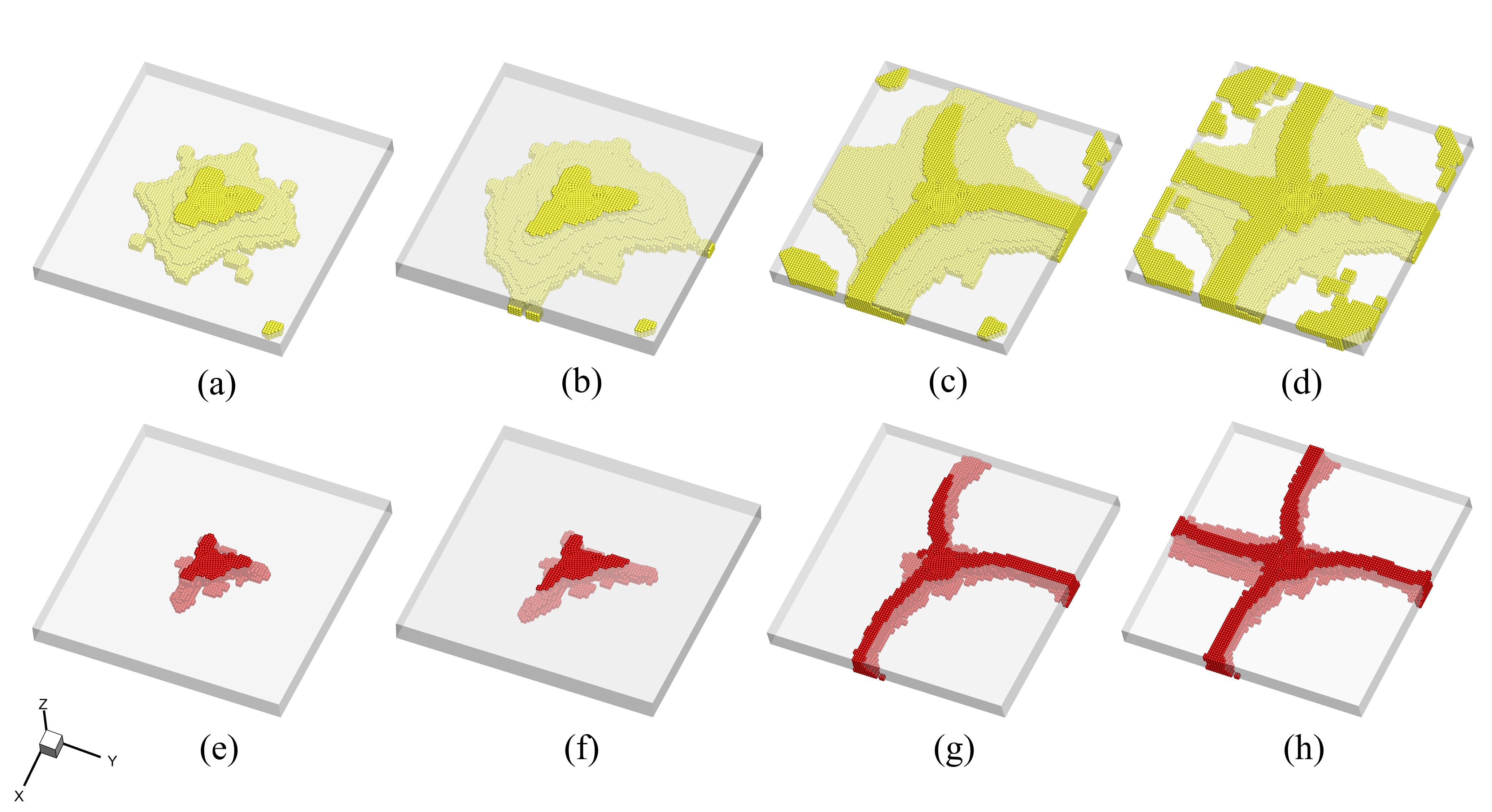}
		\caption{Top-view distributions of points with $\alpha>0$ at 14.9~ms (a), 18.7~ms (b), 22.4~ms (c), and 26.2~ms (d), and points with damage $\phi>0$ at 14.9~ms (e), 18.7~ms (f), 22.4~ms (g), and 26.2~ms (h).
		 }
		\label{3D_TOP}
	\end{figure}
 	\begin{figure}[!htbp]
		\centering
		\includegraphics[width=\textwidth]{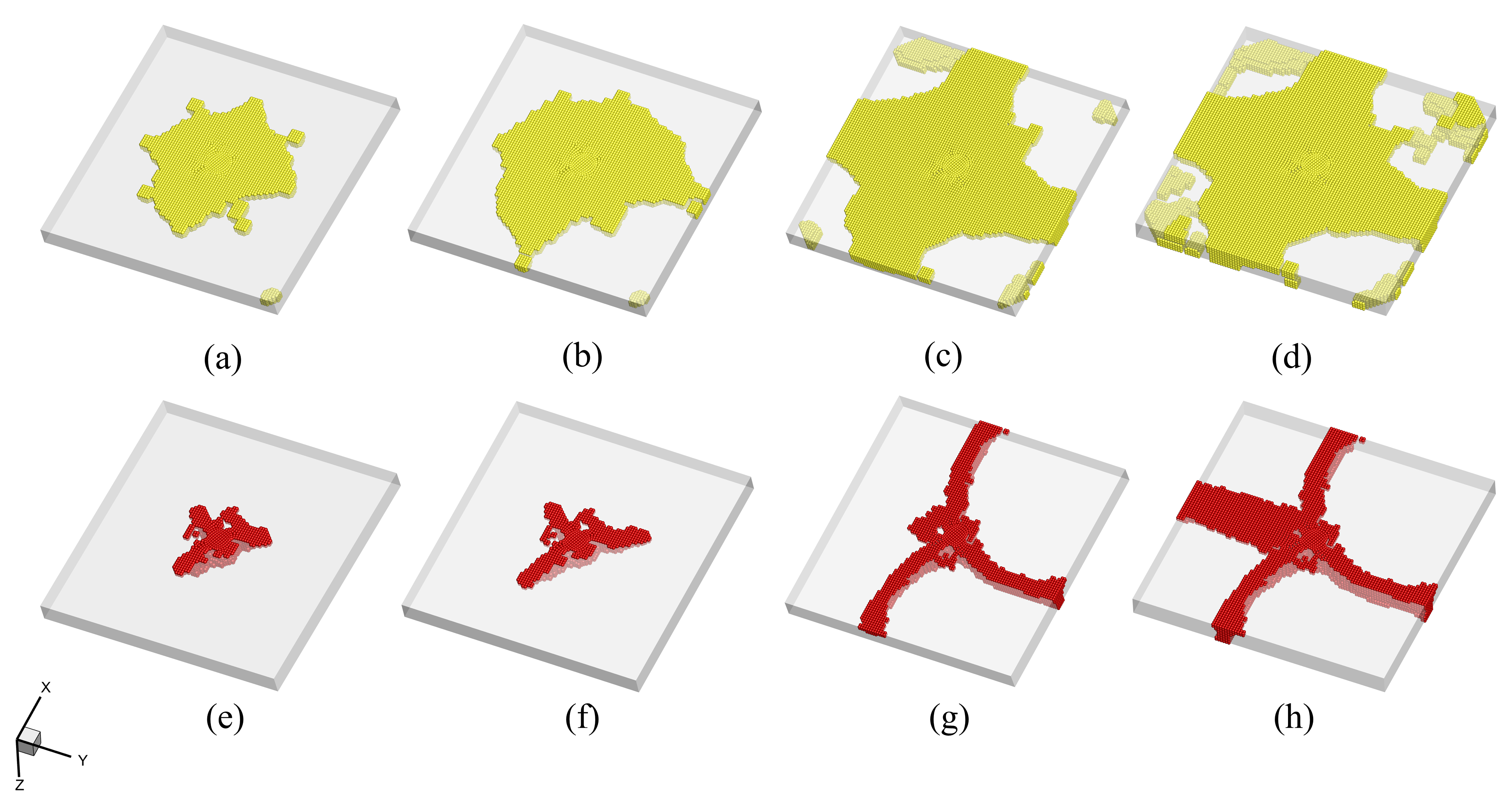}
		\caption{Bottom-view distributions of points with $\alpha>0$ at 14.9~ms (a), 18.7~ms (b), 22.4~ms (c), and 26.2~ms (d), and points with damage $\phi>0$ at 14.9~ms (e), 18.7~ms (f), 22.4~ms (g), and 26.2~ms (h).}
		\label{3D_BOT}
	\end{figure}

\begin{figure}[!htbp]
		\centering
		\includegraphics[width=\textwidth]{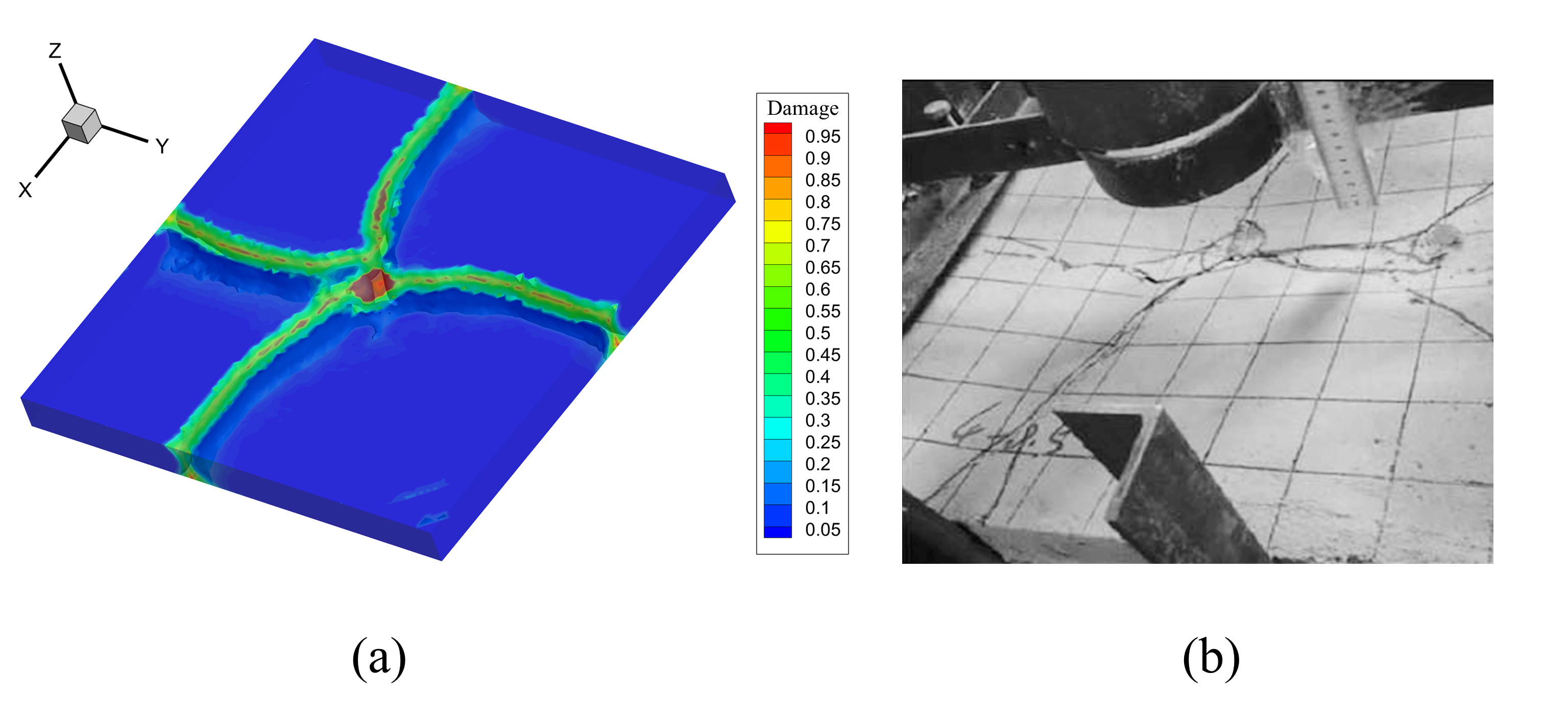}
		\caption{Comparison of crack paths obtained from the BAPD-CCM simulation (a) and experiments \cite{LOW_IMPACT} (b).}
		\label{3D_DAM_COMPARE}
	\end{figure}

    For better visualization, Figs.~\ref{3D_TOP} and \ref{3D_BOT} present the distributions of points with $\alpha>0$ and $\phi>0$ at several time instants. Fig.~\ref{3D_TOP} shows the top view, whereas Fig.~\ref{3D_BOT} shows the bottom view. The domain around the impact center is activated first and is switched to the BAPD domain, after which it gradually expands outward as damage evolves. Meanwhile, the damaged domain (i.e., points with $\phi>0$) initiates within the BAPD domain and propagates along the plate surface, progressively developing into a wider damage band and eventually forming four dominant cracks. By comparing the top and bottom views, it can also be observed that the back face (bottom surface) generally exhibits an earlier onset and a larger extent of the $\alpha>0$ domain as well as more pronounced damage, whereas the front face (top surface) mainly shows localized damage near the impact point. Furthermore, Fig.~\ref{3D_DAM_COMPARE} compares the crack paths obtained from the numerical simulation with those from the experiments. Both results show excellent agreement, which further demonstrates the effectiveness of the BAPD-CCM model in capturing dynamic fracture behavior.
    
	\section{Conclusions}\label{S8}

	This study improves the BAPD model based on the equivalence of the strain energy density, and further establishes an adaptive coupled model with CCM. The proposed correction factor can improve the solution accuracy of the BAPD model. On this basis, the proposed BAPD-CCM coupled model not only preserves the main advantages of the BAPD model in describing crack initiation and propagation, but also allows the remaining domain to be modeled by CCM, thereby improving the overall computational efficiency. Numerical results show that the proposed model can effectively handle both quasi-static and dynamic fracture problems and can provide reasonable crack evolution results.
	\section*{Acknowledgements}
	The authors gratefully acknowledge the financial support received from the Science Challenge Project, No.TZ2025001.
	\section*{Data availability}
	The data that support the findings of this study are available from the corresponding author upon reasonable request.
	\section*{CRediT authorship contribution statement}
	Wenping Han: Conceptualization, Methodology, Software, Validation, Writing -- original draft, Visualization. Bowen Sun: Validation, Investigation. Shankun Liu: Validation, Investigation. Fei Han: Conceptualization, Supervision, Project administration, Writing -- review and editing, Funding acquisition.
	\appendix
	\section[Proofs for correction factors]{Proofs for the correction factor $\chi_{\bm\xi}^{a}$}
	\label{app:chi_proofs}
	
	
	The \(H_{\bm x}\) taken as a spherical neighborhood of horizon \(\delta\) and \(\omega\) taken as a radial influence function, i.e., \(\omega=\omega(|\bm\xi|)\). Let $r=|\bm\xi|$ and $\bm n=\bm\xi/|\bm\xi|$.
	In Eq.~\eqref{CCM_E0_BA_E0_old}, the bond vector is taken as $\bm X\langle\bm\xi\rangle=\bm\xi$.
	
	The point-associated shape tensor is defined as
	\begin{equation}
		\bm K(\bm x)=\int_{H_{\bm x}}\omega(|\bm\xi|)\,\bm\xi\otimes\bm\xi\,\mathrm{d}V_{\bm\xi}.
		\label{eq:Kx_def_app}
	\end{equation}
	Using $\bm\xi=r\bm n$ and $\mathrm{d}V_{\bm\xi}=r^2\mathrm{d}r\,\mathrm{d}\Omega$ in \eqref{eq:Kx_def_app}, Eq.~\eqref{eq:Kx_def_app} can be written as
	\begin{equation}
		\bm K(\bm x)
		=\int_{0}^{\delta}\int_{S^2}\omega(r)\,r^2(\bm n\otimes\bm n)\,(r^2\mathrm{d}r\,\mathrm{d}\Omega)
		=\left(\int_{0}^{\delta}\omega(r)\,r^4\,\mathrm{d}r\right)\left(\int_{S^2}\bm n\otimes\bm n\,\mathrm{d}\Omega\right).
	\end{equation}
	By the standard spherical identity $\int_{S^2}\bm n\otimes\bm n\,\mathrm{d}\Omega=\frac{4\pi}{3}\bm I$, it follows that
	\begin{equation}
		\bm K(\bm x)=k\,\bm I,
		\qquad
		k=\frac{4\pi}{3}\int_{0}^{\delta}\omega(r)\,r^4\,\mathrm{d}r,
		\qquad
		\bm K(\bm x)^{-1}=\frac{1}{k}\bm I.
		\label{eq:Kx_kI_app}
	\end{equation}

	For each bond $\bm\xi$, the bond-associated shape tensor $\bm K_{\bm\xi}$ is defined over the domain
	$H_{\bm x}\cap h_{\bm x'}$. This domain is lens-shaped and axisymmetric about $\bm n$. The tensor $\bm K_{\bm\xi}$ can be written as
	\begin{equation}
		\bm K_{\bm\xi}=a(r)\,\bm n\otimes\bm n+b(r)\,(\bm I-\bm n\otimes\bm n),
		\label{eq:Kxi_TI_app}
	\end{equation}
	where $a(r)$ and $b(r)$ are scalar functions of $r$. Assuming $\bm K_{\bm\xi}$ is invertible, its inverse has the same form:
	\begin{equation}
		\bm K_{\bm\xi}^{-1}=p(r)\,\bm n\otimes\bm n+q(r)\,(\bm I-\bm n\otimes\bm n),
		\qquad
		p(r)=\frac{1}{a(r)},\ q(r)=\frac{1}{b(r)}.
		\label{eq:Kxi_inv_TI_app}
	\end{equation}
	
	Then we simplify the contracted term in Eq.~\eqref{CCM_E0_BA_E0_old}. Since $\bm\xi=r\bm n$ and $(\bm I-\bm n\otimes\bm n)\bm n=\bm 0$,
	\begin{equation}
		\bm K_{\bm\xi}^{-1}\bm\xi
		=\Big(p(r)\,\bm n\otimes\bm n+q(r)(\bm I-\bm n\otimes\bm n)\Big)(r\bm n)
		=p(r)\,r\,\bm n.
	\end{equation}
	And it follows that
	\begin{equation}
		\bm K_{\bm\xi}^{-1}\cdot(\bm\xi\otimes\bm\xi)
		=(\bm K_{\bm\xi}^{-1}\bm\xi)\otimes\bm\xi
		=p(r)\,r^2\,\bm n\otimes\bm n.
		\label{eq:Kxi_inv_contract_app}
	\end{equation}
	Substituting \eqref{eq:Kxi_inv_contract_app} into Eq.~\eqref{CCM_E0_BA_E0_old} gives
	\begin{equation}
		\int_{H_{\bm x}}\omega(r)\,\chi(r)\,p(r)\,r^2\,\bm n\otimes\bm n\,\mathrm{d}V_{\bm\xi}
		=\bm I.
		\label{eq:Eq39_reduced_tensor_app}
	\end{equation}
	Transforming to spherical coordinates gives
	\begin{equation}
		\left(\int_{0}^{\delta}\omega(r)\,\chi(r)\,p(r)\,r^4\,\mathrm{d}r\right)
		\left(\int_{S^2}\bm n\otimes\bm n\,\mathrm{d}\Omega\right)
		=\bm I.
	\end{equation}
	Thus using $\int_{S^2}\bm n\otimes\bm n\,\mathrm{d}\Omega=\frac{4\pi}{3}\bm I$, Eq.~\eqref{CCM_E0_BA_E0_old} is equivalent to the scalar condition
	\begin{equation}
		\int_{0}^{\delta}\omega(r)\,\chi(r)\,p(r)\,r^4\,\mathrm{d}r=\frac{3}{4\pi}.
		\label{eq:scalar_condition_app}
	\end{equation}
	
	A new correction factor $\chi_{\bm\xi}^{a}$ is considered
	\begin{equation}
		\chi_{\bm\xi}^{a}=\frac{\bm n^{T}\bm K(\bm x)^{-1}\bm n}{\bm n^{T}\bm K_{\bm\xi}^{-1}\bm n}.
		\label{eq:chi_a_def_app}
	\end{equation}
	Using \eqref{eq:Kx_kI_app} gives $\bm n^{T}\bm K(\bm x)^{-1}\bm n=\frac{1}{k}$, while
	\eqref{eq:Kxi_inv_TI_app} implies $\bm n^{T}\bm K_{\bm\xi}^{-1}\bm n=p(r)$. Hence
	\begin{equation}
		\chi_{\bm\xi}^{a} = \chi^{a}(r)=\frac{1}{k\,p(r)}.
		\label{eq:chi_a_closed_app}
	\end{equation}
	Substituting \eqref{eq:chi_a_closed_app} into \eqref{eq:scalar_condition_app} yields
	\begin{equation}
		\int_{0}^{\delta}\omega(r)\,\chi^{a}(r)\,p(r)\,r^4\,\mathrm{d}r
		=\frac{1}{k}\int_{0}^{\delta}\omega(r)\,r^4\,\mathrm{d}r
		=\frac{1}{k}\cdot\frac{3k}{4\pi}
		=\frac{3}{4\pi},
	\end{equation}
	where \eqref{eq:Kx_kI_app} was used in the second equality. Therefore, $\chi_{\bm\xi}^{a}$ satisfies Eq.~\eqref{CCM_E0_BA_E0_old} exactly.
	
	\clearpage
	\bibliographystyle{unsrt}      
	\bibliography{bibtex}   
\end{document}